\documentclass[12pt,twocolumn]{aastex63}

\newcommand{\kms}{\rm km~s^{-1}}

\newcommand{\dn}{D_{n}4000}

\usepackage{color}
\usepackage{multirow}
\usepackage{amsmath}
\usepackage{appendix}
\usepackage{hhline}
\usepackage{graphicx}

\begin{document}

\title {The HectoMAP Redshift Survey: First Data Release}

\author{Jubee Sohn}
\affiliation{Smithsonian Astrophysical Observatory, 60 Garden Street, Cambridge, MA 02138, USA}

\author{Margaret J. Geller}
\affiliation{Smithsonian Astrophysical Observatory, 60 Garden Street, Cambridge, MA 02138, USA}

\author{Ho Seong Hwang}
\affiliation{Korea Astronomy and Space Science Institute, 776 Daedeokdae-ro, Yuseong-gu, Daejeon 34055, Korea}

\author{Daniel G. Fabricant}
\affiliation{Smithsonian Astrophysical Observatory, 60 Garden Street, Cambridge, MA 02138, USA}

\author{Sean M. Moran}
\affiliation{Smithsonian Astrophysical Observatory, 60 Garden Street, Cambridge, MA 02138, USA}

\author{Yousuke Utsumi}
\affiliation{Kavli Institute for Particle Astrophysics and Cosmology, SLAC National Accelerator Laboratory, Stanford University, 2575 Sand Hill Road, Menlo Park, CA 94025, USA}

\email{jubee.sohn@cfa.harvard.edu}

\begin{abstract}
HectoMAP is a dense, red-selected redshift survey to a limiting $r = 21.3$ covering 55 square degrees in a contiguous 1.5$^\circ$ strip across the northern sky. This region is also covered by the Subaru/Hyper Suprime-Cam (HSC) Subaru Strategic Program (SSP) photometric survey enabling a range of applications that combine a dense foreground redshift survey with both strong and weak lensing maps. The median redshift of HectoMAP exceeds 0.3 throughout the survey region and the mean density of the redshift survey is $\sim 2000$ galaxies deg$^{-2}$. Here we report a total of 17,313 redshifts in a first data release covering 8.7 square degrees. We include the derived quantities $\dn$ and stellar mass for nearly all of the objects. Among these galaxies, 8117 constitute a 79\% complete red-selected subsample with $r \leq 20.5$ and an additional 4318 constitute a 68\% complete red-selected subsample with $20.5 < r < 21.3$. As examples of the strengths of HectoMAP data we discuss two applications: refined membership of redMaPPer photometrically selected clusters and a test of HSC photometric redshifts. We highlight a remarkable redMaPPer strong lensing system. The comparison of photometric redshifts with spectroscopic redshifts in a dense survey uncovers subtle systematic issues in the photometric redshifts.
\end{abstract}

\section{Introduction}

Redshift surveys are a pillar of modern cosmology. They chart the way galaxies mark the distribution of matter in the universe on large scales and they hold clues to the way individual galaxies and systems of galaxies evolve. 

Beginning in the 1980's digital detector technology enabled advances in the extent and quality of redshift surveys (e.g. \citealp{Davis82,Geller89,Shectman96}). Over the last twenty-five years, wide-field multi-object spectrographs have enabled a fantastic array of surveys that cover large regions of the universe at relatively  recent epochs (e.g. \citealp{Strauss02, Colless03, Ahn14, Alam15, Liske15, Drinkwater18}) and probe the universe at redshifts $ z > 1$ (e.g. \citealp{Lilly09, Newman13, Silverman15, Scodeggio18}). The next decade promises even more spectacular advances with more heavily multiplexed systems (e.g. \citealp{Tamura18, Finoguenov19, Maiolino20}).

Hectospec, a 300-fiber wide-field instrument on the 6.5-meter MMT, has played an important role in redshift surveys of galaxies covering the most recent 7 Gigayears of the history of the universe \citep{Kochanek12, Geller14, Geller16, Zahid16}. The SHELS survey \citep{Geller05, Geller10} was the first to combine a foreground dense redshift survey with a weak lensing map to explore the combined power of these modern cosmological tools (see also \citealp{Geller14, Hwang16}).

The SHELS survey \citep{Geller16}, complete to an extinction corrected $R = 20.2$ over roughly 8 square degrees, also provides a benchmark for more extensive color-selected surveys. An MMT survey of the COSMOS field, hCOSMOS, provides additional photometric and spectroscopic calibration \citep{Damjanov18}.

HectoMAP, carried out with Hectospec on the MMT, is a red-selected redshift survey covering 55 square degrees to a limiting $r = 21.3$ in a narrow strip across the northern sky. The Subaru/Hyper Suprime-Cam (HSC) Subaru Strategic Program (SSP) \citep{Aihara18} also covers the HectoMAP region with deep 5-band photometry. Here we describe the first data release (HectoMAP DR1) for the HectoMAP survey covering 8.7 square degrees of the survey. This region of the HectoMAP DR1 includes the region covered by the HSC SSP public DR1.

The average density of HectoMAP is $\sim~2000$ galaxies per square degree. This high density is the signature property of the redshift survey that underlies its central scientific goals \citep{Geller15}. These goals include identification and characterization of clusters (e.g. \citealp{Sohn18a, Sohn18b}) and voids \citep{Hwang16} throughout the survey region. Because massive clusters accrete roughly half of their mass from a redshift of 0.5 to the present, the HectoMAP survey serves as a baseline for dynamical measures of cluster accretion history.

The powerful combination of HSC SSP imaging and HectoMAP is a foundation for exploring the way galaxies trace the large-scale matter distribution by combining a weak lensing map with the foreground redshift survey. The combined data also provide a host of strong lensing candidate systems from individual massive galaxies to massive clusters of galaxies.

The HectoMAP survey fills an interesting niche for enhanced understanding of the evolution of the quiescent galaxy population and its dependence on environment. \citet{Damjanov19} use SHELS data for 4,200 quiescent objects to investigate relationships among stellar mass, size, and stellar population age to redshift $z \sim 0.6$. HectoMAP DR1 contains nearly 12,000 quiescent objects and the entire HectoMAP survey will contain $\sim 70,000$ for more extensive investigations in this redshift range. The density, completeness, and deep HSC photometry of the final HectoMAP sample will enable insights into the dependence of quiescent galaxy size evolution on environment (e.g. \citealp{Damjanov15, Gargiulo19}). 

We provide a redshift, $\dn$ and stellar mass for a total of 17,313 galaxies in the 8.7 square degree HectoMAP DR1 region. Amogn these, there are  9,775 galaxies are in highly complete red subsamples of galaxies with $r < 21.3$. Section \ref{SDATA} includes discussion of the SDSS and Subaru HSC photometry, the spectroscopy, the derived parameters $\dn$ and stellar mass, the completeness of the survey, and the redshift distribution. We highlight two direct applications in Section \ref{Sec:apps}. Section \ref{Sec:cluster} extends our earlier investigations of redMaPPer clusters; Section \ref{Sec:photz} demonstrates the power of HectoMAP for investigation of subtle systematics in photometric redshifts. We use cosmological parameters $\Omega_{m} = 0.3$, $\Omega_{\Lambda} = 0.7$, and H$_{0}$ = 70 km/s/Mpc throughout.

\section{The Data}\label{SDATA}

HectoMAP is a dense redshift survey with a median redshift $\sim 0.3$. The average number density of galaxies with spectroscopic redshifts is $\sim~2000$ deg$^{-2}$ ($\sim1200$ galaxies deg$^{-2}$ within the highly complete red-selected subsample \citep{Geller11, Geller15, Hwang16, Sohn18a, Sohn18b}). The selection for the complete HectoMAP sample is $(g-r)_{model,0} > 1$ for $r_{petro,0} \leq  20.5$; for $20.5 < r_{petro,0} \leq 21.3$ we include an additional constraint, $(r-i)_{model,0} > 0.5$. Here, $r_{petro, 0}$ refers to the SDSS Petrosian magnitude  corrected for foreground extinction. The SDSS model colors, $(g-r)_{model, 0}$ and $(r-i)_{model, 0}$, are also corrected for foreground extinction. Hereafter we designate $r = r_{petro,0}$, $(g-r) = (g-r)_{model,0}$, and $(r-i) = (r-i)_{model,0}$ in the text; we retain the full notation in the figure captions for clarity.

The full HectoMAP survey covers 54.64 deg$^2$ within the boundaries 200 $<$ R.A. (deg) $<$ 250 and 42.5 $<$ Decl. (deg) $<$ 44.0. Here we include redshifts, stellar masses and $\dn$ for galaxies in the region 242 $<$ R.A. (deg) $<$ 250 and 42.5 $<$ Decl. (deg) $<$ 44.0, an area of 8.7 square degrees that includes the HSC SSP  public DR1 region \citep{Aihara18}. The median redshift for HectoMAP in this data release is $z = 0.31$. Hereafter we refer to the this data release as HectoMAP DR1.

In 2009 we observed the GTO2deg$^{2}$ field \citep{Miyazaki07} as a test of the feasibility of the HectoMAP project. We acquired 4405 redshifts (out of 4541) with Hectospec covering this 2.1 deg$^2$ field \citep{Kurtz12} which overlaps slightly with HectoMAP DR1. The original selection for red objects in this region was $(r-i) > 0.4$ throughout the apparent magnitude range, $r < 21.3$. In the full HectoMAP survey we have no $(r-i)$ cut for galaxies with $r \leq 20.5$ and a tighter $(r - i) > 0.5$ for objects with apparent magnitudes between 20.5 and 21.3.

We reduced the Hectospec spectroscopy in the GTO2deg$^{2}$ field \citep{Kurtz12} with an IRAF based pipeline \citep{Kurtz98} including the RVSAO cross-correlation package \citep{Kurtz98}. We have rereduced these data with the current pipeline, HSRED v2.0; we include the rereduced results for the 505 galaxies in the small overlap region here. There is essentially no difference between the redshifts returned by the two pipelines
We discuss the overall HectoMAP sample selection based on SDSS photometry (Section \ref{Sec:SDSS}), the available HSC photometry (Section \ref{Sec:HSC}), and the MMT Hectospec spectroscopy (Section \ref{Sec:Hectospec}). We discuss the completeness of the survey and the redshift distribution in Section \ref{Sec:comp}. Sections \ref{Sec:D4000} and \ref{Sec:mass} include descriptions of the parameters we derive from the spectroscopy: $\dn$ and stellar mass.

\subsection{SDSS Photometry}\label{Sec:SDSS}

The SDSS provides the photometric basis for HectoMAP. Because observations for HectoMAP extend over 10 years, we used the SDSS data releases beginning with DR7 \citep{SDSSDR7} and progressing to DR9 \citep{SDSSDR9}. We have now updated all of the catalogs to SDSS DR16 \citep{SDSSDR16}, the basis for all of the quantities quoted here.

The primary HectoMAP survey targets are red galaxies with $(g - r) > 1.0$ for $r \leq 20.5$ (the bright red subsample) and galaxies with $20.5 < r \leq  21.3$, $(g - r) > 1$ and $(r- i) > 0.5$ (the faint red subsample). We select galaxies based on the SDSS $probPSF = 0$, where {\it probPSF} is the probability that the object is a star. The color cut for the faint red sample removes the galaxies with $z \lesssim 0.2$; the cut is necessary because the observing time required without it is prohibitive. For both the bright and faint subsamples we remove low surface brightness objects that are generally beyond the limits of the MMT spectroscopy by requiring $r_{fiber,0} < 22.0$. Table \ref{number} lists the number of photometric objects in the entire sample and in various subsamples.

\begin{deluxetable*}{lcccccc}
\tablecaption{Number of Galaxies in Subsamples\label{number}}
\tablecolumns{7}
\tabletypesize{\scriptsize}
\tablewidth{0pt}
\tablehead{
\colhead{Subsample} & \colhead{$N_{phot}$} & 
\colhead{$N_{spec, Hecto}^{1}$}
& \colhead{$N_{spec, SDSS}^{2}$}  & 
\colhead{$N_{spec, NED}^{3}$}  & \colhead{$N_{spec, total}^{4}$} & 
\colhead{Completeness (\%)}} 
\startdata
Entire Sample ($r \leq 23.0$)                    		& \nodata &	14767	&	2409	&	137	& 17313	& \nodata \\
Supplementary ($21.3 < r \leq 23.0$)            & \nodata &	1083		&	384	&	0		& 1467		& \nodata \\
\hhline{=======}
Main ($r \leq 21.3$)                             			& 41006	&	13684	&	2025	&	137	& 15846	& 38.6 \\
Main ($g-r > 1.0, r \leq 21.3$)                  		& 20526	&	11509	&	1441	&	43		& 12993	& 63.3 \\
{\bf Main ($g-r > 1.0, r-i > 0.5, r \leq 21.3$)}	& 12164	&	8561		&	1211	&	3		& 9775		& {\bf 80.4} \\
Main ($g-r \leq 1.0$)                            			& 20480	&	2175		&	584	&	94		& 2853		& 13.9 \\
\hline
Bright ($r \leq 20.5$)                           			& 20076	&	9015		&	1590	&	135	& 10740	& 53.5 \\
{\bf Bright ($g-r > 1.0$)}                       		& 10305	&	7051		&	1023	&	43		& 8117		& {\bf 78.8} \\
{\bf Bright ($g-r > 1.0, r-i > 0.5$)}            		& 5837   	&	4660		&	794	&	3		& 5457		& {\bf 93.5} \\
Bright ($g-r \leq 1.0$)                          			& 9771   	&	1964		&	567	&	92		& 2623		& 26.8 \\
\hline
Faint                                            					& 20930	&	4669		&	435	&	2		& 5106		& 24.4			\\
Faint  ($g-r > 1.0$)                             			& 10221	&	4458		&	418	&	0		& 4876		& 47.7 			\\
{\bf Faint  ($g-r > 1.0, r-i > 0.5$)}            		& 6327		&	3901		&	417	&	0		& 4318		& {\bf 68.2}	\\
Faint  ($g-r \leq 1.0$)                          			& 10709	&	211		&	17		&	2		& 230		&  2.1
\enddata
\tablenotetext{1}{Number of spectroscopic redshifts from MMT/Hectospec.}
\tablenotetext{2}{Number of spectroscopic redshifts from SDSS and BOSS.} 
\tablenotetext{3}{Number of spectroscopic redshifts from NED including \citet{Gronwall04, Jaffe13}. }
\tablenotetext{4}{Total number of spectroscopic redshifts.}
\end{deluxetable*}

\subsection{HSC Photometry}\label{Sec:HSC}

\begin{figure*}
\centering
\includegraphics[scale=0.47]{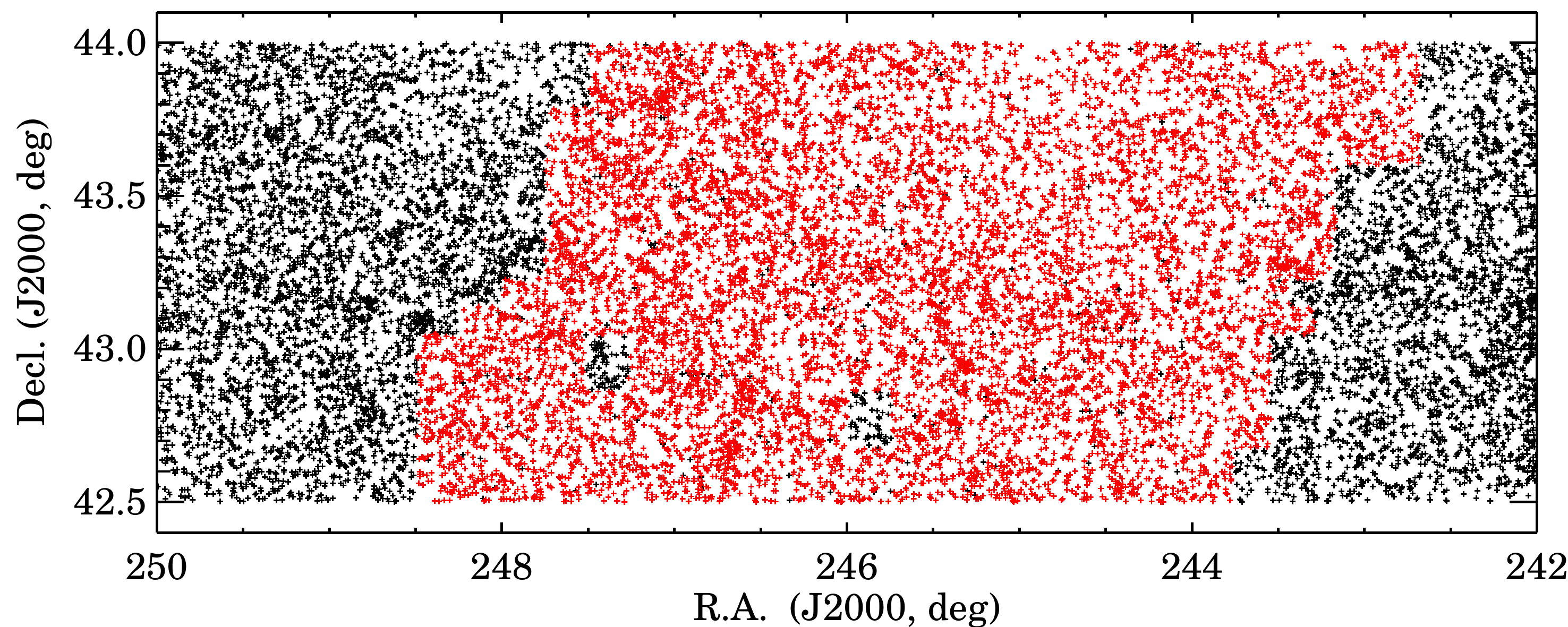}
\caption{HectoMAP DR1 footprint. 
Black points are galaxies with spectra in the HectoMAP DR1. 
Red points are galaxies with photometric counterparts in the HSC SSP DR1 catalog. 
Note the two significant islands of black within the red patch where  HSC data are missing. } 
\label{footprint}
\end{figure*}

The HectoMAP region is included in the Wide portion of the Subaru/HSC \citep{Miyazaki12} Subaru Strategic Program (SSP). The first public HSC SSP data release \citep{Aihara18} includes 4.7 square degrees of the HectoMAP region. Figure \ref{footprint} shows the HectoMAP DR1 region, 242 $<$ R.A. (deg) $<$ 250 and 42.5 $<$ Decl. (deg) $<$ 44.0, along with the 4.7 square degree footprint of the included HSC SSP DR1 (red points). The total area covered by the HectoMAP release is 8.7 square degrees. 

The HSC Wide photometry includes data in $g$, $r$, $i$, $z$, and $y$ to $5\sigma$ depths of 26.8, 26.4, 26.2, 25.4, and 24.7, respectively, for point sources. Because our main goal is to provide spectroscopic data and because our observations were based on SDSS photometry, we do not use the HSC photometry explicitly here. 

There are two sizeable missing patches (black islands in Figure \ref{footprint}) in the HectoMAP HSC DR1 region in all bands. \citet{Aihara18} also highlight several general issues that affect the relatively bright ($i < 19$) galaxies in HectoMAP. For example, some galaxies in the HectoMAP magnitude range are shredded and most galaxies with $i < 19$ have composite Model (cModel) magnitudes inconsistent with the SDSS.

We take advantage of the public release of photometric redshifts in the HSC SSP public DR2 \citep{Aihara19}. Here, we use HSC SSP DR2 because it provides photometric redshifts for  galaxies in the entire HectoMAP DR1 region. \citet{Tanaka18} derive photometric redshifts based on a variety of techniques ranging from classical template-fitting to machine learning techniques. They test these techniques against available spectroscopy. Although the test samples used by \citet{Tanaka18} are large, they emphasize the need for additional dense, complete, independent spectroscopic data. HectoMAP DR1 is much shallower than the HSC SSP data, but the density and completeness of the spectroscopy over its depth and area are unique. We discuss the HSC photometric redshifts in Section \ref{Sec:photz}.

\subsection{Hectospec spectroscopy}\label{Sec:Hectospec}

We measured redshifts with the 300-fiber Hectospec mounted on the MMT 6.5-meter telescope \citep{Fabricant05} from 2009 to 2019. Hectospec deploys 300 fibers over a 1 square degree field-of-view and covers the wavelength range of 3700-9100~\AA. The standard total exposure time for a HectoMAP observation is one hour in three 20 minute segments to enable cosmic ray removal. In good conditions a single Hectospec observation returns $\sim 250$ redshifts. The typical yield is $\sim 200$ reliable redshifts. 

Obtaining a uniformly complete redshift survey over the large HectoMAP field is a major observational challenge. To position Hectospec for each run, we first evaluated the spectroscopic completeness of the survey at the beginning of the run in $0.25 \times 0.25$ degree pixels. We then ranked the pixels from the least to most complete. Finally we chose a set of Hectospec positions that maximized coverage of the highly ranked (most incomplete) pixels. Variable observing conditions affected the yield for each position. Over the ten year survey we revisited each position in the entire survey $\sim 10$ times. In the DR1 region, the typical number of revisits is $\sim 9$.

It is not always possible to fill all of the Hectospec fibers with survey targets. We used remaining fibers to observe bluer objects prioritized by apparent magnitude (brighter objects without a redshift had higher priority). This approach yielded 2205 Hectospec redshifts for unique objects bluer than the survey selection limits (Table \ref{number}).

\begin{figure*}
\centering
\includegraphics[scale=0.7]{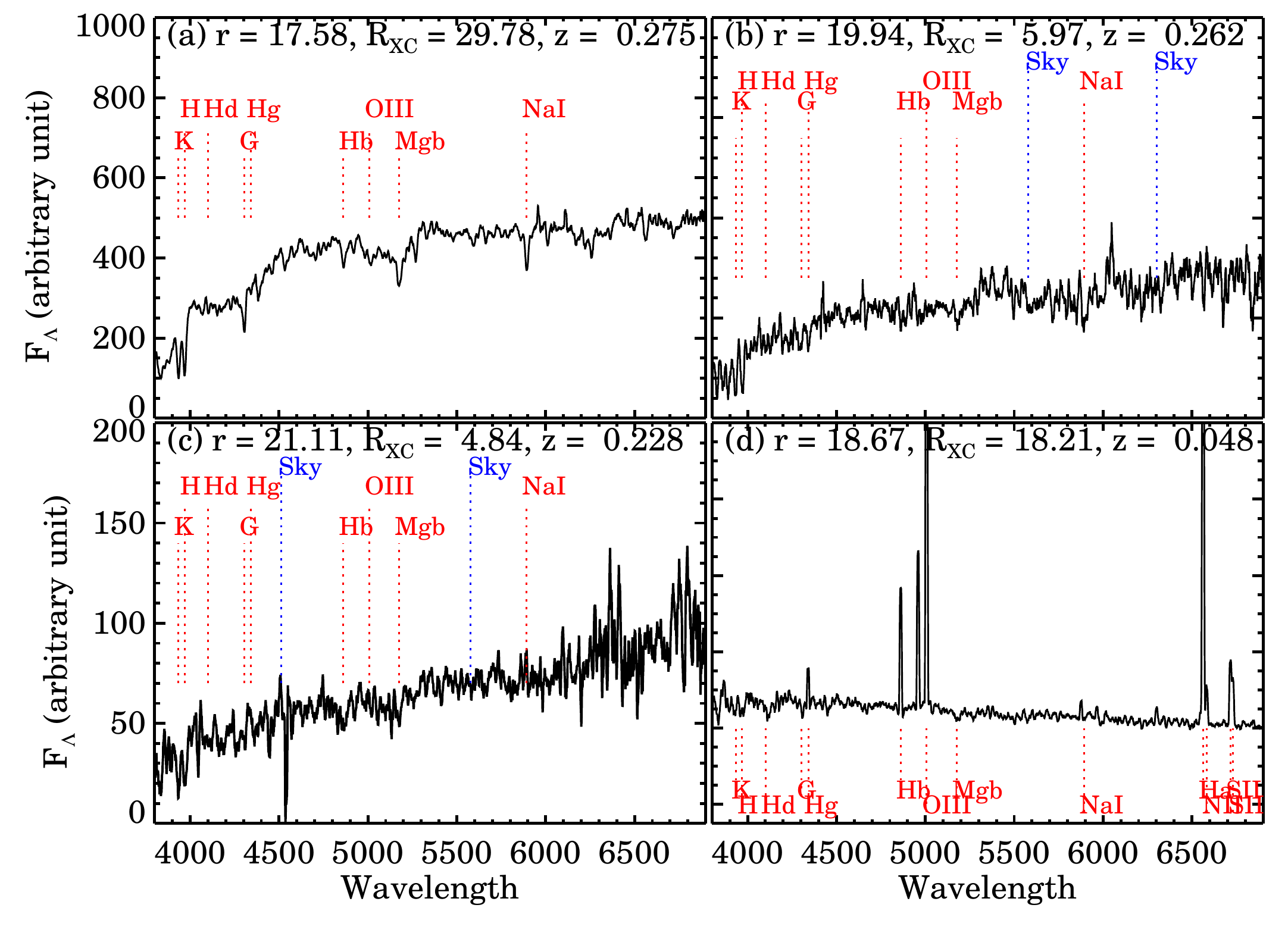}
\caption{Sample spectra for  galaxies in the HectoMAP DR1 region. } 
\label{spectra}
\end{figure*}

Figure \ref{spectra} shows three absorption line spectra (panels a-c) for galaxies at $z \sim 0.25$ covering the range of the cross-correlation quality measure, $R_{XC}$ \citep{Tonry79}. In all cases the H and K break is apparent. These absorption-line objects are the main focus of HectoMAP. In the lower right panel (d) we show an emission line spectrum at lower redshift; the object has negligible continuum and strong lines.

We show the spectra for a fixed window in the rest frame. The variation in the quality of the spectra is driven largely by observing conditions including lunar phase (for the bright portion of the survey), seeing, and transparency. The pipeline provides a standard indicator of the quality of the spectrum, $R_{XC}$, a measure of the significance of the cross-correlation peak \citep{Tonry79}.

\begin{figure}
\centering
\includegraphics[scale=0.47]{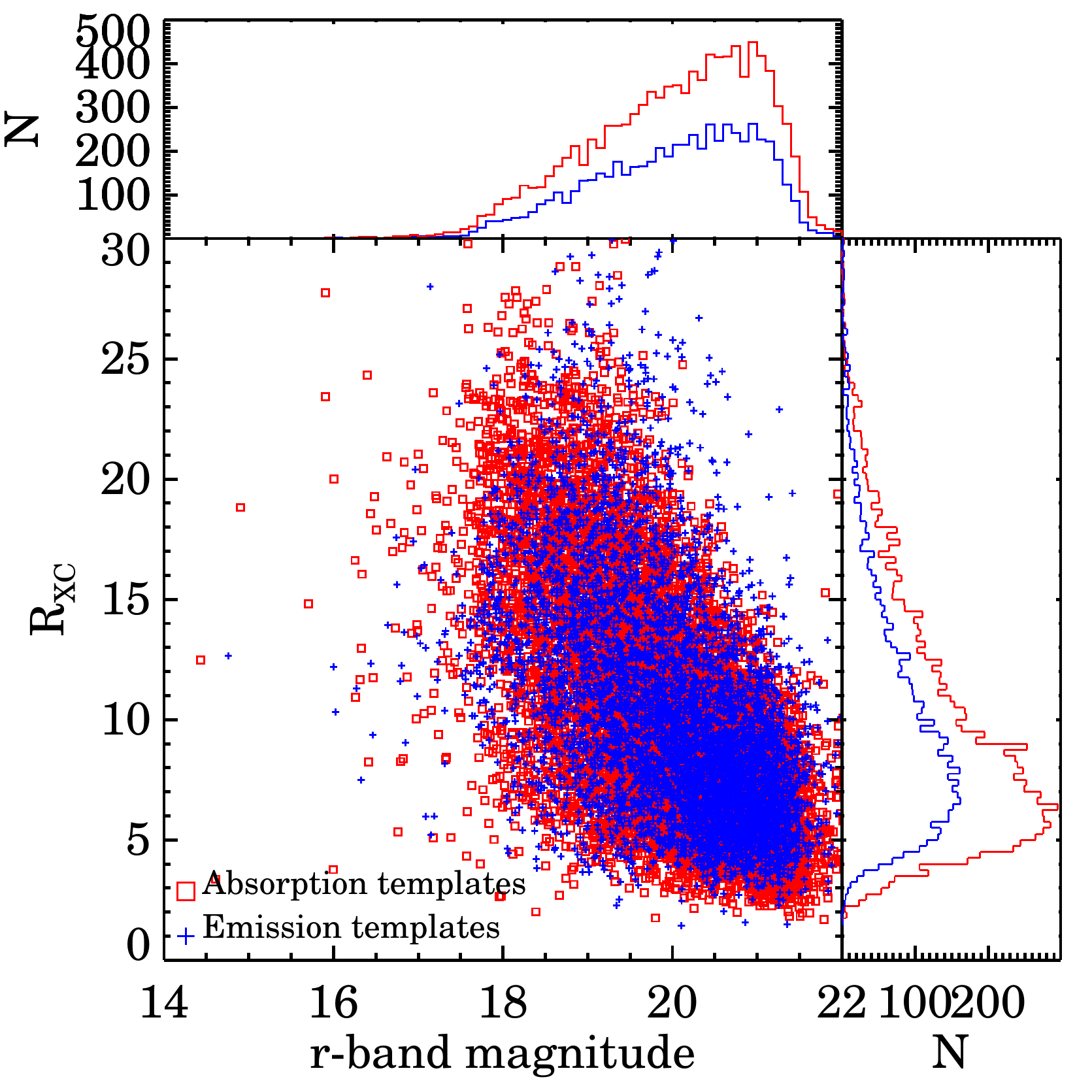}
\caption{$R_{XC}$ from RVSAO as a function of $r_{petro,0}$. 
Red and blue points show redshifts derived from emission- and absorption-line spectral templates, respectively.
The histograms display $r_{petro,0}$ and $R_{XC}$ distributions for emission (blue) and absorption(red)-line redshifts. } 
\label{czxcr}
\end{figure}

Figure \ref{czxcr} shows $R_{XC}$ as a function of extinction corrected $r$. We show the distribution of $R_{XC}$ for galaxies with a redshift derived from an absorption-line template (red points) and for galaxies with redshifts derived from an emission-line template (blue points). The emission-line templates are overrepresented among the brightest and faintest galaxies and among galaxies with the highest and lowest $R_{XC}$. At bright apparent magnitudes we sample relatively more blue objects; at the faintest magnitudes, objects with strong emission lines are more likely to yield a reliable redshift (e.g. \citealp{Glazebrook2007}). At the lowest $R_{XC}$, we identify objects with a reliable redshift by visual inspection. The presence of more than one emission line makes the visual identification more likely. The highest $R_{XC}$s reflect the presence of strong emission lines (see the example in Figure \ref{spectra}).

Although we could use $R_{XC}$ alone to select reliable redshifts, we inspect each spectrum visually after the pipeline processing. Visual inspection recovers or discards (1) spectra with badly subtracted night sky lines, (2) spectra where there are two objects at different redshifts in the same fiber aperture and (3) AGN and quasars which do not match the templates. On the basis of the visual inspection we classify objects: `Q' for high quality, `?' for doubtful cases, and `X' for complete failures. We conservatively report a redshift only for a spectrum classified as Q.

Table \ref{number} lists the number of photometric objects in various color and magnitude ranges (main, bright red, faint red, blue). For each sample in the Table, we list the number of unique Hectospec spectra with Qs (column 3) and the additional number of unique objects from the combination of SDSS and BOSS in each sample (column 4). Column 5 lists the  number of reliable redshifts in each subsample. In some cases this number exceeds the sum of columns 3 and 4 because we include a small number of redshifts from the NASA Extragalactic Database (NED, including \citealp{Gronwall04, Jaffe13}). We highlight (bold-face) the sample and subsamples where we strive for completeness within the magnitude and color limits. Although we select extended sources with $probPSF = 0$, 524 objects turn out to be stars. In other words, among the objects with Q quality spectra, $\sim 3\%$ are actually stars listed  in Appendix A. Thus the estimates of completeness based on the sets of photometric SDSS extended objects (Table \ref{number}) are probably slight underestimates because some fraction of the remaining objects are also stars.

Table \ref{hmdata} compiles the 17313 HectoMAP DR1 redshifts. Among the galaxies in Table \ref{hmdata} 1467 are fainter than $r = 21.3$. These faint objects account for the excess relative to the 15846 objects in Table \ref{number} that reviews the main survey targets. We include the SDSS ID (column 1), the right ascension and declination (columns 2 and 3) the SDSS $r$, $(g-r)$, and $(r-i)$ (columns 4 -6), the redshift with its error (column 7), the source of the redshift (column 8), the stellar mass and $\dn$ (columns 9 and 10), and a redMaPPer cluster membership indicator (column 11). The error in the redshift in column 8 is the formal error returned by cross-correlation. The median formal error is $28~\kms$; ($24~\kms$ and $31~\kms$ for emission and absorption line spectra, respectively). 

An earlier Hectospec redshift survey, SHELS, provides a measure of the internal error based on a set of repeat measurements intended for this purpose \citep{Geller14}. SHELS is comprised of two fields, F1 and F2. In the F2 field a set of repeat measurements for 1651 unique objects provides the internal estimate (normalized by $(1 + z)$); for emission line objects the internal error is $24~\kms$ and for absorption line objects (the bulk of the HectoMAP sample) the internal error is $48~\kms$. In their Hectospec survey of the COSMOS field based on the same observing protocols followed for HectoMAP, \citet{Damjanov19} obtain similar internal errors for emission and absorption line redshifts, $26~\kms$ and $42~\kms$, respectively. It is interesting that these internal errors are essentially identical to the XCSAO estimate for emission line objects but they exceed it by about 50\% for absorption line spectra. 

\begin{figure}
\centering
\includegraphics[scale=0.5]{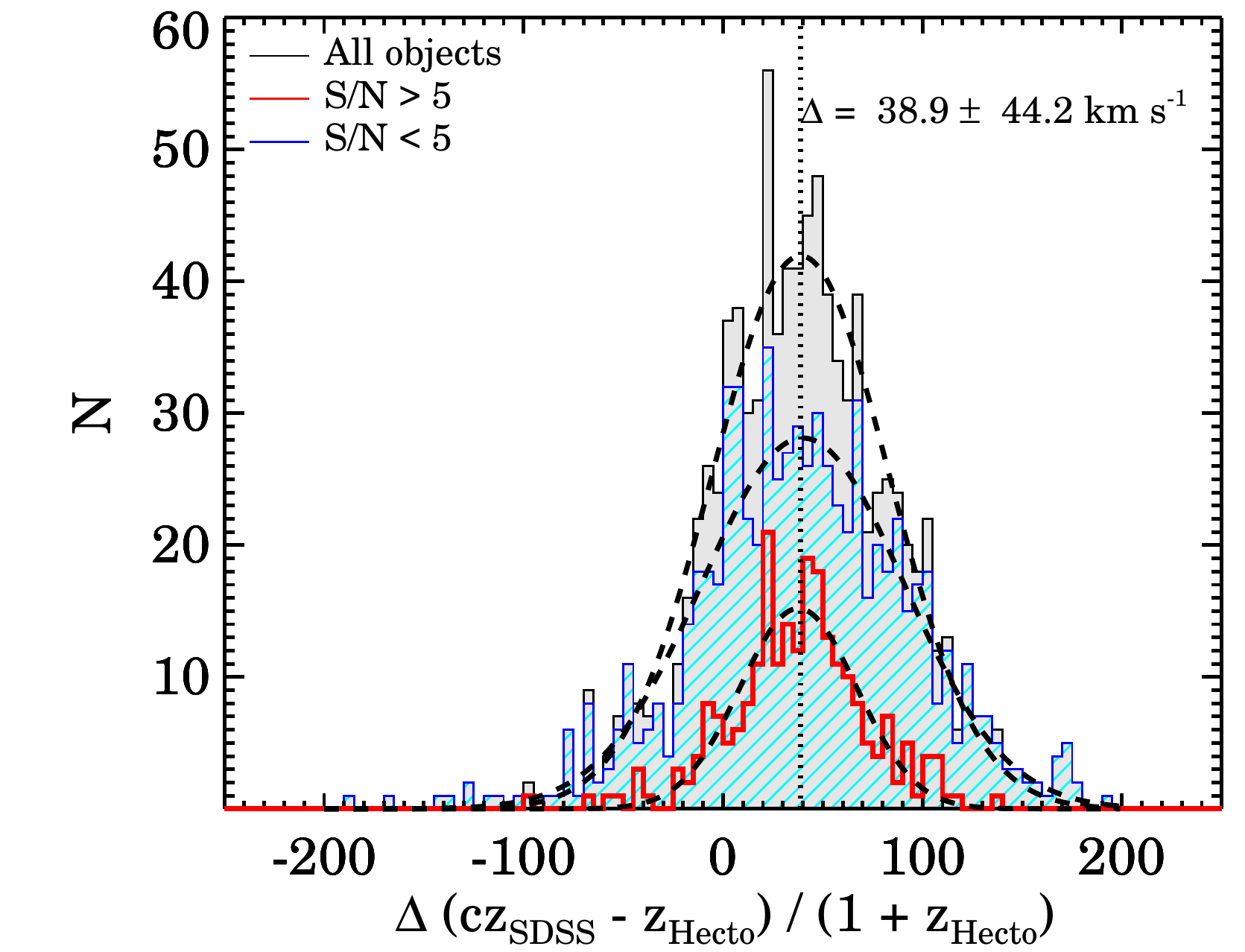}
\caption{Distributions of redshift differences between Hectospec and SDSS/BOSS  for galaxies with duplicate measurements in the HectoMAP DR1 region. The black filled histogram shows all objects with both Hectospec and SDSS/BOSS redshifts. Red open and blue hatched histograms show spectra with high signal to noise ($S/N > 5$) or  low S/N , respectively. The dashed curves show the best-fit Gaussians for each distribution. The vertical dotted line indicates the Gaussian mean for the entire sample (black histogram). } 
\label{zcomp}
\end{figure}

As a measure of the external error, Figure \ref{zcomp} examines the HectoMAP DR1 overlap with SDSS/BOSS. There are 129 and 852 objects with Hectospec observations that overlap SDSS and BOSS observations, respectively. The black histogram in Figure \ref{zcomp} shows the distribution of redshift differences ($\Delta c(z_{SDSS/BOSS} - z_{Hecto}) / (1 + z_{Hecto})$) between Hectospec and SDSS (or BOSS) for all overlaps. We derive the best-fit Gaussian for this distribution (dashed curve). The mean and standard deviation for the total overlapping sample are $39~\kms$ and $44~\kms$, respectively. There is a small systematic offset of $39~\kms$ between Hectospec and SDSS/BOSS redshifts in the sense that Hectospec redshifts are slightly larger, but this offset is comparable with the $1\sigma$ standard deviation, $44~\kms$. We note that the offset is somewhat smaller at low redshift: $\Delta c(z_{SDSS/BOSS} - z_{Hecto} / (1 + z_{Hecto})) = 26 \pm 32~\kms$ at $z < 0.15$. However, the fraction of HectoMAP DR1 galaxies with $z < 0.15$ is only $\sim 10\%$. 

We also plot redshift differences for overlapping objects with high (red histogram) and low (blue histogram) signal to noise (S/N) Hectospec spectra in Figure \ref{zcomp}. Both distributions are well represented by Gaussians. The best-fit Gaussian mean and standard deviation for the high S/N and low S/N overlapping samples are $38 \pm 29~\kms$ and $40 \pm 51~\kms$, respectively. The mean redshift offset between Hectospec and SDSS/BOSS is insensitive to the S/N of the Hectospec spectra; the external error increases with decreasing S/N as expected. \citet{Damjanov18} use 2661 overlapping objects to find that Hectospec redshifts exceed their zCOSMOS counterparts by a small zero-point offset of $17\pm 2~\kms$.

The small systematic offsets between HectoMAP and SDSS/BOSS and between Hectospec and zCOSMOS redshifts are comparable with or less than the $1\sigma$ error in HectoMAP redshifts based on the sum in quadrature of the internal and external errors. In fact, for the entire HectoMAP sample, the external error alone (Figure \ref{zcomp}) slightly exceeds the systematic shift. We used both XCSAO and template fitting applied to the data and to a set of simulated spectra to try to identify the source of these systematics, but the tests failed to reveal the underlying, fundamental cause. Because these small systematics are irrelevant for most applications of the survey, we simply summarize them here for completeness.

\begin{deluxetable*}{lcccccccccc}
\tablecaption{HectoMAP DR1 Catalog\label{hmdata}}
\tablecolumns{11}
\tabletypesize{\scriptsize}
\tablewidth{0pt}
\tablehead{
\colhead{SDSS Object ID} & \colhead{R.A.} & \colhead{Decl.} & \colhead{$r_{petro, 0}$} & \colhead{$(g-r)^{*}$} & \colhead{$(r-i)^{*}$} & \colhead{z} & \colhead{z Source} & \colhead{$M_{*}$} & \colhead{$\dn$} & \colhead{redMaPPer} }
\startdata
1237659326027596446 & 243.530137 & 42.506127 & $20.56 \pm  0.11$ &  1.66 &  0.54 & $0.31258 \pm 0.00029$ & MMT  & $ 10.48^{+0.10}_{-0.16}$ & $2.17 \pm 0.24$ &   \nodata \\
1237659326564205123 & 243.530863 & 43.202909 & $20.00 \pm  0.17$ &  0.81 &  0.09 & $0.13372 \pm 0.00002$ & MMT  & $  9.02^{+0.17}_{-0.19}$ & $1.12 \pm 0.03$ &   \nodata \\ 
1237659327100945547 & 243.531558 & 43.757552 & $21.33 \pm  0.08$ &  1.98 &  1.01 & $0.74919 \pm 0.00026$ & BOSS & $ 11.33^{+0.18}_{-0.19}$ & $1.54 \pm 0.09$ &   \nodata \\
1237659327100945058 & 243.531600 & 43.778505 & $20.53 \pm  0.04$ &  1.56 &  0.78 & $0.49998 \pm 0.00021$ & MMT  & $ 11.14^{+0.10}_{-0.10}$ & $1.52 \pm 0.05$ &   \nodata \\
1237659326564205607 & 243.531601 & 43.259037 & $20.42 \pm  0.13$ &  0.94 &  0.45 & $0.25664 \pm 0.00012$ & MMT  & $ 10.23^{+0.14}_{-0.16}$ & $1.67 \pm 0.05$ & HMRM05629
\enddata
\tablenotetext{*}{Foreground extinction-corrected model colors. }
\end{deluxetable*}

\subsection{Survey Completeness and Redshift Distribution}\label{Sec:comp}

\begin{figure}
\centering
\includegraphics[scale=0.5]{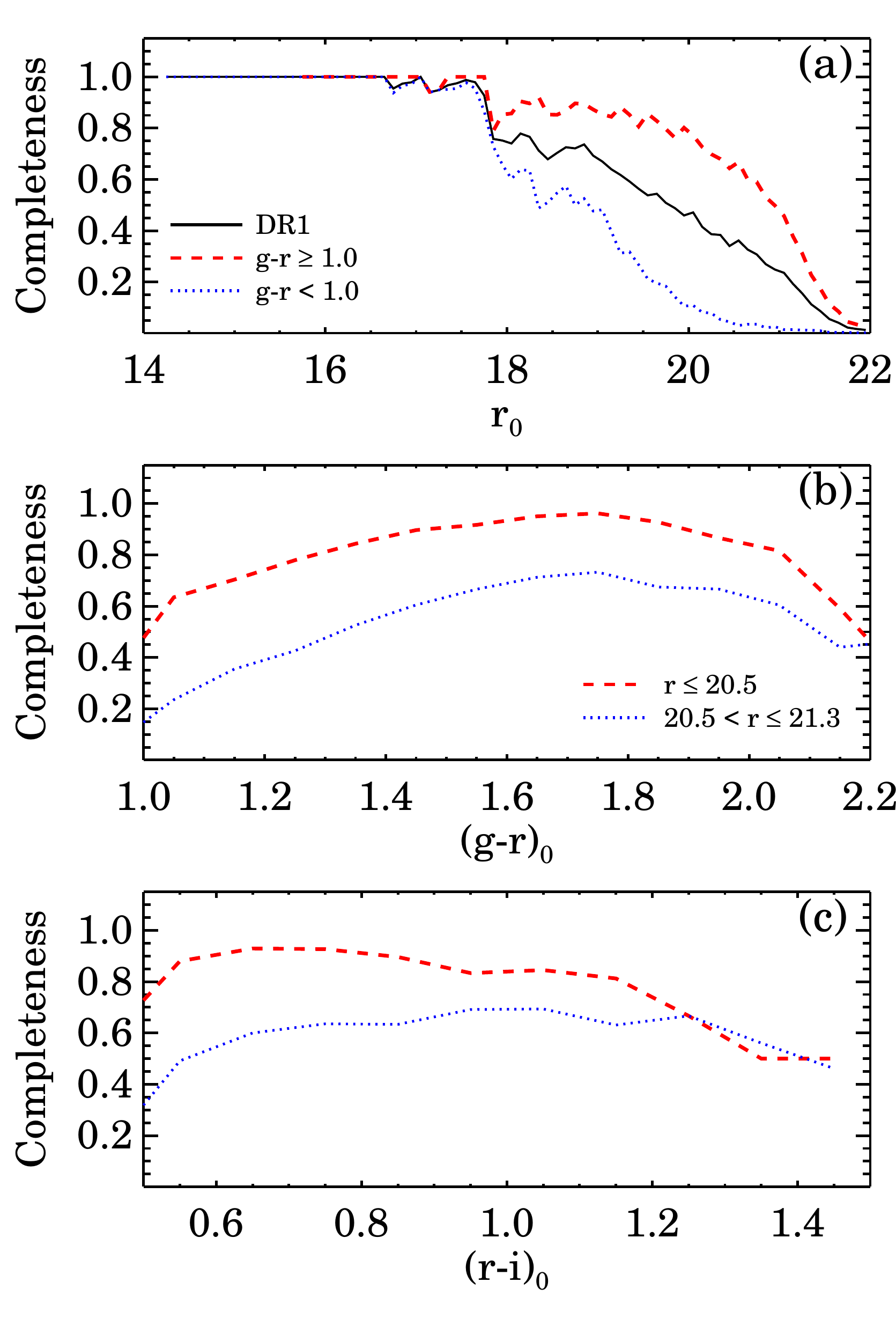}
\caption{
(a) Spectroscopic survey differential completeness as a function of $r_{petro,0}$. 
Solid, dashed, and dotted lines show the completeness for  all, red ($(g-r)_{model,0}\geq 1.0$) and blue ($(g-r)_{model,0}< 1.0$) galaxies.
(b) Survey completeness as a function of $(g-r)_{model,0}$  for the bright ($r_{petro,0} \leq 20.5$) and faint ($20.5 < r_{petro,0} \leq 21.3$) subsamples. 
(c) Same as (b), but as a function of $(r-i)_{model,0}$ . } 
\label{complete_mag}
\end{figure}

The upper panel of Figure \ref{complete_mag} shows the differential completeness of the redshift survey as a function of $r$ for the entire DR1 and for subsamples segregated by $(g-r)$ (see Table \ref{number}). We made no attempt to obtain a complete blue sample ($(g-r) < 1$). We show the curve merely to indicate the content of the dataset. The differential completeness of the red-selected $(g-r) \geq 1$ subsample exceeds $\sim 75\%$ for $r < 20$ and drops to 50\% for $r < 20.9$. 

Dense redshift surveys that overlap the redshift range of HectoMAP include SDSS/BOSS \citep{Ahn14, Alam15}, GAMA \citep{Liske15}, and SHELS \citep{Geller16}. The SDSS Main Galaxy sample \citep{Strauss02} is $\gtrsim 94\%$ complete to a limiting $r = 17.77$ over a large region; there is no color selection. BOSS \citep{Ahn14, Alam15} covers the magnitude range, but it is color-selected and sparse (see Figure \ref{cone} below). GAMA \citep{Liske15} covers about 5.5 times the area of HectoMAP to a limiting $r = 19.8$ in an equatorial region. The completeness of GAMA generally exceeds 94\% and there is no color selection. Finally, SHELS covers $\sim 8$ square degrees with a completeness  $\gtrsim 94$\% to $R \sim 20.2$ ($r \sim 20.5$) in two well-separated fields. The depth is comparable with the bright portion of HectoMAP and there is no color selection. The set of overlapping surveys with no color selection provide a basis for assessing the impact on scientific results resulting from the red color-selection in HectoMAP.

Figures \ref{complete_mag} (b) and (c) show the fractional completeness of the $r \leq 20.5$ (red dashed line) and the $20.5 < r \leq 21.3$ (blue dotted line) subsamples of HectoMAP as a function of $(g-r)$ (a) and $(r-i)$ (b). For the faint subsample the completeness declines for lower values of $(g-r)$ as a result of the cut in $(r-i)$. In the brighter portion with $r \leq 20.5$ the completeness also declines for bluer $(g-r)$ because we initially prioritized galaxies with $(r-i) > 0.5$ \citep{Hwang16,Sohn18a,Sohn18b}. In general the completeness declines for the reddest galaxies because they are at redshifts $\gtrsim 0.5$ and thus tend to be faint, lower surface brightness objects (see Figure 1 of \citealp{Hwang16}). 

\begin{figure*}
\centering
\includegraphics[scale=0.5]{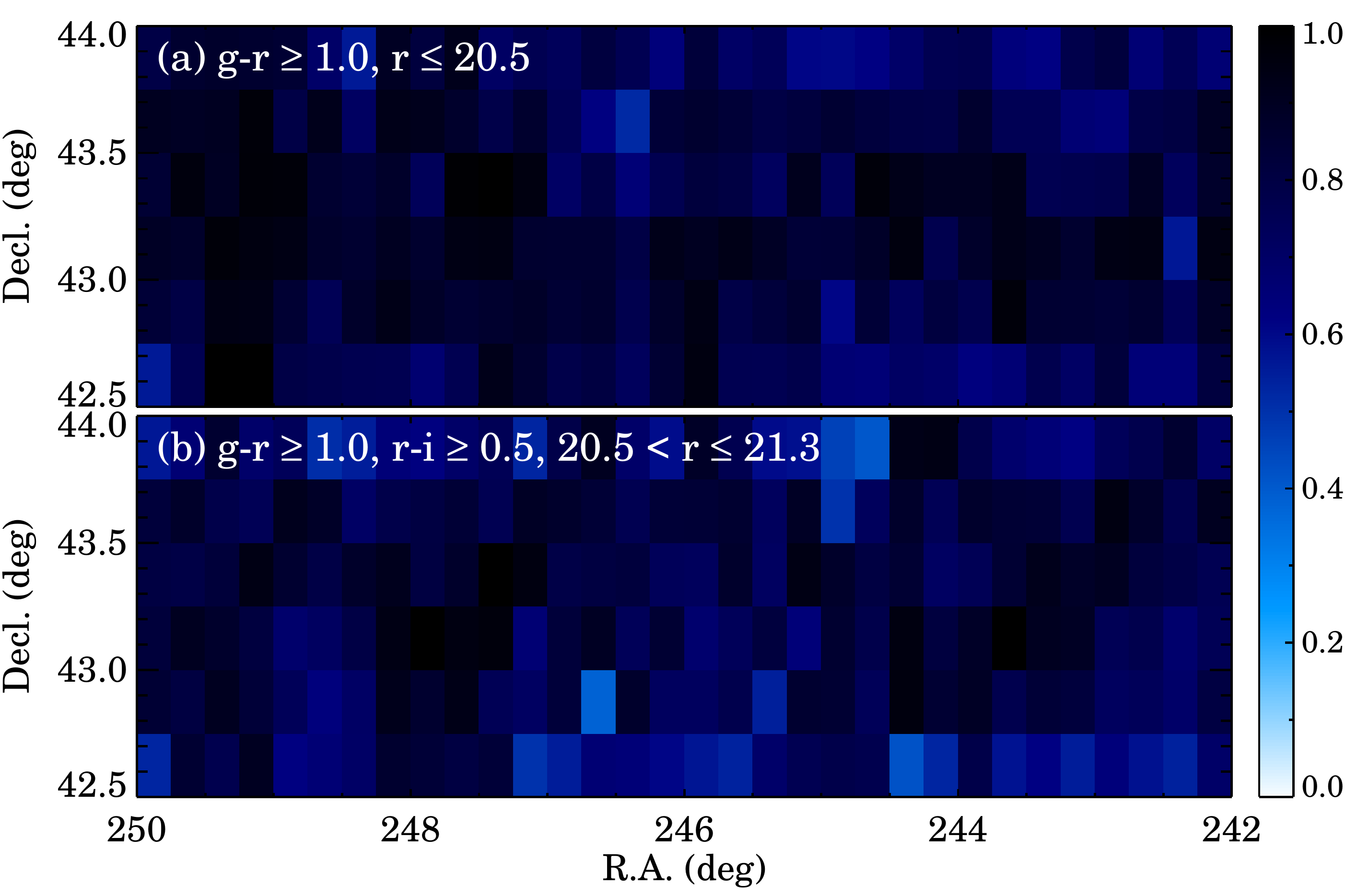}
\caption{
(a) Survey completeness in $0.25^\circ {\rm x}0.25^\circ$ pixels for red-selected galaxies with 
 $r_{petro,0} \leq 20.5$ in HectoMAP DR1.  
Darker colors indicate more complete pixels.
(b) Survey completeness as in (a) but for red-selected objects with $20.5 < r_{petro,0} < 21.3$  } 
\label{complete_map}
\end{figure*}

Figure \ref{complete_map} (a) shows the completeness of the red-selected sample with $r \leq 20.5$ as a function of position on the sky. The bins are $0.25^\circ \times 0.25^\circ$. The completeness is remarkably uniform ($> 80\%$). Only 15\% of the pixels are less than 70\% complete; there are no pixels with a completeness $< 50\%$.

Figure \ref{complete_map} (b) shows the completeness on the sky for red-selected galaxies with $20.5 < r \leq 21.3$. For this magnitude range there is an $(r-i)$ selection in addition to the constraint $(g-r) \geq 1$. Only 29\% of the pixels are less than 70\% complete; 5\% of the pixels have a completeness less than 50\%. The fainter portion of HectoMAP is less complete because of the greater sensitivity  to observing conditions. In particular, the observations are more sensitive to  seeing. The survey is also less complete near the declination boundaries where observations are inefficient.

\begin{figure}
\centering
\includegraphics[scale=0.5]{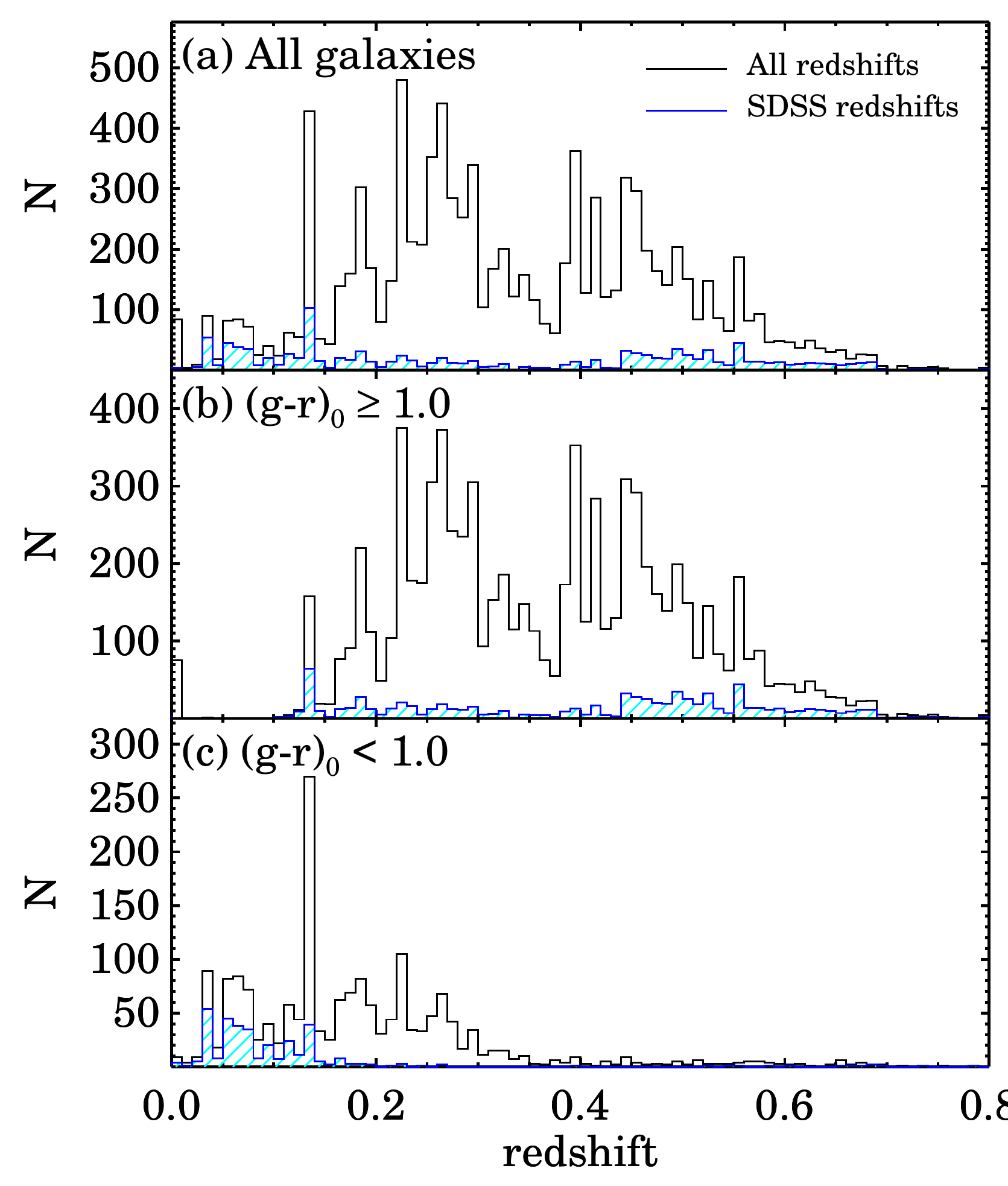}
\caption{Redshift distribution for HectoMAP DR1 (open histogram):
 (a) all galaxies, (b) $(g-r)_{model,0} \geq 1.0$ and (c) $(g-r)_{model,0} < 1.0$. 
Blue hatched histograms show the corresponding SDSS/BOSS redshift distributions.} 
\label{zhist}
\end{figure}

Figure \ref{zhist} shows the redshift distribution for the HectoMAP DR1 sample and for two subsamples segregated by $(g-r)$. In each panel the blue hashed histogram shows redshifts from SDSS/BOSS. From the upper to lower panel, the median redshifts are $\sim 0.31$, $\sim 0.36$, and $\sim 0.16$, respectively. 

The sharply defined peaks are the signature of the large-scale structure of the universe. It is striking that the peaks appear clustered over broad redshift ranges. The low density region in the redshift range $\sim 0.30 - 0.38$ is remarkable. The spatial scale corresponding to this deficit is an impressive $\sim 290$ Mpc in the radial direction. This scale is comparable with the largest HectoMAP void found by \citet{Hwang16} (see their Figure 16). Voids with this extent occupy a long tail in the N-body simulations explored by \citet{Hwang16}.

Figure \ref{cone} contrasts the dense sampling of the HectoMAP survey (right) with the combined SDSS/BOSS (left). In the redshift range $0.1 < z < 0.8$, there are 17202 galaxies in the HectoMAP cone and 2518 in the SDSS/BOSS display. The HectoMAP survey clearly delineates voids, filaments, walls, and clusters (as a guide, redMaPPer clusters are indicated by red circles in each panel); these structures are barely visible in the SDSS/BOSS display.

\citet{Sutter14} emphasize the importance of dense sampling for the definition of void sizes, shapes, and density profiles. \citet{Hwang16} compare a volume limited subsample of the HectoMAP data with the 300 mock surveys drawn from the Horizon Run 4 N-body simulation to show the efficacy of HectoMAP in defining the elements of the large-scale structure in the universe. The data and the models are in excellent agreement. The full HectoMAP sample will provide a test bed for void evolution for $z \lesssim 0.7$.

HectoMAP has already provided a test \citep{Sohn18a} of the photometric redMaPPer cluster catalog \citep{Rykoff14, Rykoff16} over the redshift range $0.08 < z < 0.6$. About 90\% of the redMaPPer systems in the HectoMAP footprint correspond to systems in the HectoMAP survey. This test included even low richness redMaPPer systems. On average HectoMAP provides a spectroscopic redshift for $\sim 20$ cluster members. At redshift $z > 0.35$, the HectoMAP spectroscopic richness does not correlate well with the redMaPPer richness, but the HectoMAP target selection could be an issue \citep{Sohn18a}. We reconsider the redMaPPer systems in HectoMAP DR1 in Section \ref{Sec:cluster} below.

\begin{figure*}
\centering
\includegraphics[scale=0.32]{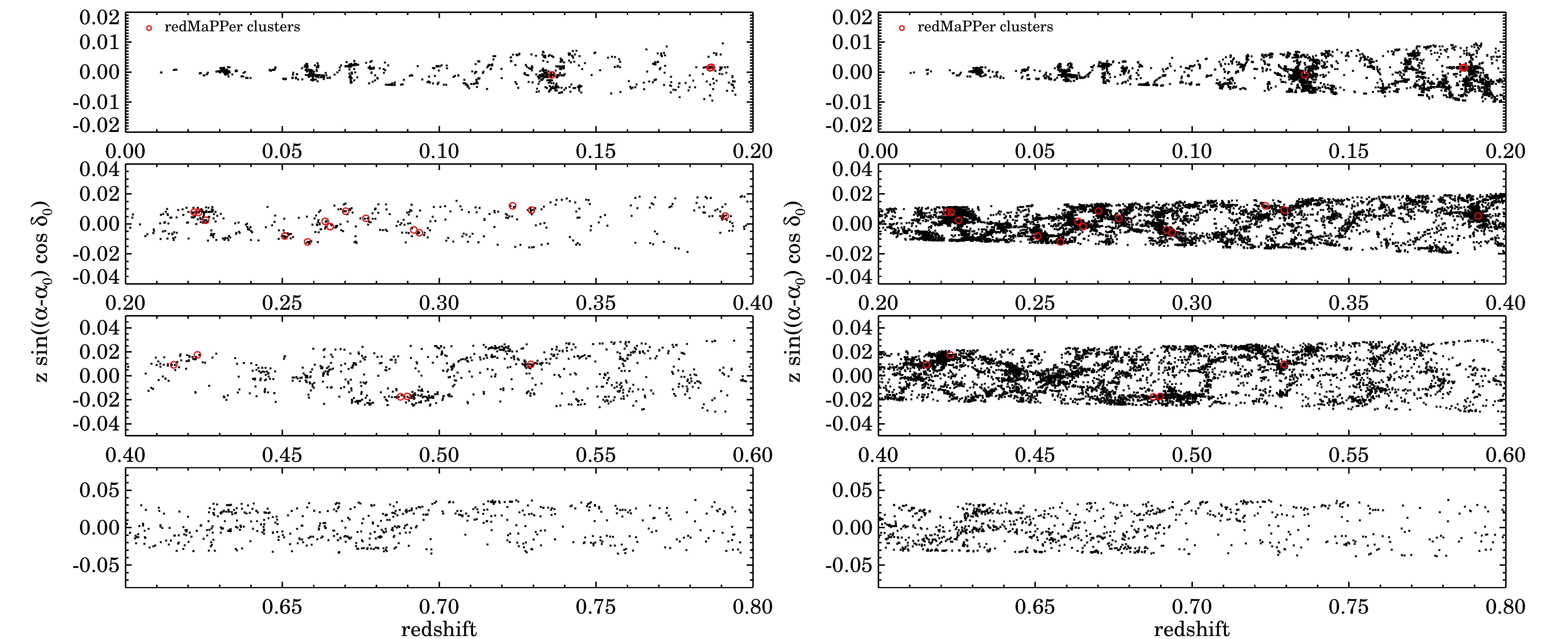}
\caption{Cone diagrams for  HectoMAP DR1: (left) galaxies from SDSS/BOSS and (right) HectoMAP. 
Red circles indicate redMaPPer clusters (Section \ref{Sec:cluster}). } 
\label{cone}
\end{figure*}

\subsection{$\dn$}\label{Sec:D4000}

We compute the stellar population age sensitive indicator $\dn$. This spectral index is the flux ratio between two spectral windows (3850 - 3950 \AA\ and 4000 - 4100 \AA) near the 4000\AA\ break \citep{Balogh99}. \citet{Fabricant08} show that the typical error in $\dn$ is 0.045 times the value of the index and for spectra overlapping with the SDSS, agreement is excellent. We include a value of $\dn$ for more than 99\% of the objects in both the bright and faint red-selected subsamples of HectoMAP (Table \ref{mcomp}). For 1\% of objects, we could not measure $\dn$ because the redshifts come from the NED. 

\begin{figure*}
\centering
\includegraphics[scale=0.45]{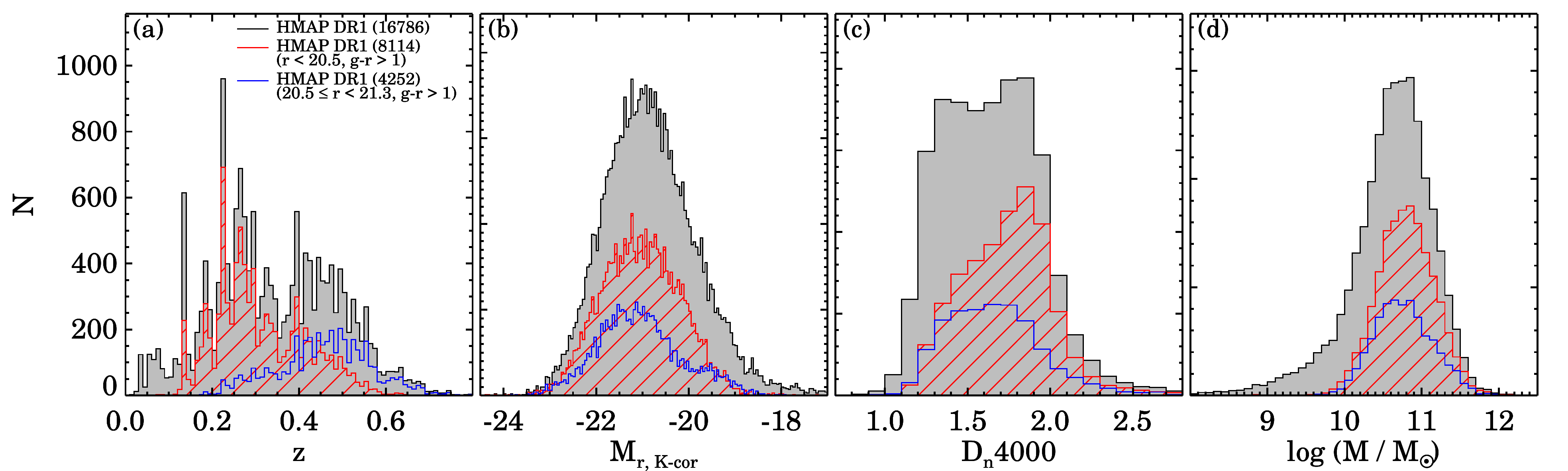}
\caption{Distributions of HectoMAP DR1 galaxy properties: (a) redshift, (b) K-corrected absolute magnitude, (c) $\dn$, (d) stellar mass. 
Black shaded histograms show all HectoMAP DR1 galaxies.
Red hatched and blue open histograms display galaxies in the bright, red subsample ($r < 20.5$ and $(g-r)_{model,0} > 1.0$) and in the faint, red subsample ($20.5 \leq r_{petro,0} < 21.3$, $(g-r)_{model,0} > 1.0$, $(r-i)_{model,0} > 0.5$), respectively. }
\label{param}
\end{figure*}

Panels (a) and (b) of Figure \ref{param} show the redshift and absolute magnitude distributions for the entire HectoMAP DR1 sample (black), the bright red-selected sample (red hatched histogram), and the faint red-selected sample (blue). The faint sample includes mostly galaxies at $z > 0.4$ with a tail of intrinsically lower luminosity (lower stellar mass) objects toward lower redshift; these objects are evident in panel (d). The bright red selected sample peaks around the $z \sim 0.3$ with a significant extension of intrinsically luminous galaxies at $z > 0.4$. These fiducial distributions underlie the distributions of other derived physical properties in panels (c) and (d).

Figure \ref{param} (c) shows the $\dn$ distribution for all of the galaxies in HectoMAP DR1 (open histogram). The red hatched histogram shows the $\dn$ distribution for the bright red-selected subsample with $r < 20.5$ and $(g-r) > 1.0$. The red color selection moves the $\dn$ distribution for bright red-selected sample toward higher $\dn$ as expected. The blue histogram shows the $\dn$ for the faint red selected sample. The peak shifts toward lower $\dn$, an indication of the younger stellar age of the galaxies.

Table \ref{mcomp} lists the fraction of galaxies with $\dn > 1.5$ in each subsample. We use this cut as a proxy for identifying the quiescent population (e.g., \citet{Moresco13}, \citet{Damjanov18}). Because of the red selection, quiescent galaxies dominate HectoMAP DR1. For the bright sample, the fraction of $\dn > 1.5$ galaxies exceeds that for the fainter sample primarily as a result of the combination of the fixed observed color cuts and the lower median redshift of the subsample.

\begin{deluxetable*}{lcccc}
\tablecaption{Completeness of Stellar Mass and $\dn$ Measurements\label{mcomp}}
\tablecolumns{5}
\tabletypesize{\scriptsize}
\tablewidth{0pt}
\tablehead{
\colhead{Subsample} & \colhead{$N_{spec}$} & 
\colhead{$f_{\dn}^{1, *}$}& \colhead{$f_{M*}^{2, \dagger}$} & 
\colhead{$f_{\dn > 1.5}^{3}$}}
\startdata
All											&	17313	&	99.0	&	99.9		& 64.3 \\
$g-r > 1, r < 20.5$				&	8117		&	99.3	&	99.9		& 78.3 \\
$g-r > 1, 20.5 \leq r < 21.3$	&	4876		&	99.8	&	100.0	& 64.6
\enddata
\tablenotetext{1}{$f_{\dn}$ is the fraction of galaxies with $\dn$ measurements. }
\tablenotetext{2}{$f_{M*}$ is the fraction of galaxies with stellar mass estimates.}
\tablenotetext{3}{$f_{\dn > 1.5}$ is the fraction of galaxies with $\dn > 1.5$ among the galaxies with spectroscopic redshifts.}
\tablenotetext{*}{We could not measure }
\end{deluxetable*}

\subsection{Stellar Mass}\label{Sec:mass}

For consistency with previous MMT redshift surveys (e.g., \citealp{Geller14, Zahid16}), we calculate stellar masses based on SDSS $ugriz$ model magnitudes corrected for foreground extinction. We fit the observed spectral energy distribution (SED) with the Le PHARE fitting code \citep{Arnouts99, Ilbert06}. We use the stellar population synthesis models of \citet{Bruzual03} and we assume a universal Chabrier initial mass function (IMF, \citealp{Chabrier03}). We consider a suite of models with two metallicities and with exponentially declining star formation rates. The e-folding times for the star formation ranges from 0.1 to 30 Gyr. Model SEDs include various extinction levels and stellar population ages. We explore the internal extinction range $E (B-V) = 0 - 0.6$ based on the \citet{Calzetti00} extinction law. The population age range is 0.01 to 13 Gyr. We normalize each SED to solar luminosity. The ratio between the observed and synthetic SED is the stellar mass. We take the median of the distribution of best fit stellar masses as the estimate of the stellar mass for a particular object. 

Figure \ref{param} (d) shows the stellar mass distribution for HectoMAP DR1. Essentially all of the galaxies in HectoMAP DR1 have a stellar mass estimate (Table \ref{mcomp}); the few objects without stellar mass estimates have inconsistent photometry (particularly in $u$-band).  The stellar mass distribution for the bright red-selected subsample (red hashed histogram) contains a relatively larger fraction of massive objects than the entire DR1 sample (black histogram). Bluer objects with lower stellar mass dominate the low stellar mass tail of the black histogram; for these objects HectoMAP is not complete. In contrast, the red hashed histogram represents a highly complete subsample (Table \ref{number}).

\section{Two Examples of HectoMAP Applications}\label{Sec:apps}

The HectoMAP survey has broad applications including computation of statistical measures of large-scale structure and comparison of the large-scale matter distribution as traced by galaxies and by weak lensing. We plan to address these and other issues in future work.

Here, we highlight two straightforward applications of HectoMAP: details of photometrically identified redMaPPer clusters of galaxies as revealed by spectroscopy, and a test of HSC photometric redshifts at bright magnitudes. The dense sampling of HectoMAP is designed for examining the properties and evolution of clusters of galaxies at redshifts from 0.2 to 0.7. Section \ref{Sec:cluster} uses the current larger HectoMAP dataset to enhance the results of \citet{Sohn18a} and \citet{Sohn18b}. Section \ref{Sec:photz} tests the power of HectoMAP as a probe of the accuracy of HSC SSP photometric redshifts. These applications highlight both the strengths (e.g. density, depth, and completeness) and limitations (e.g. color selection) of the HectoMAP survey.

\subsection{redMaPPer Clusters}\label{Sec:cluster}

One of the central goals of HectoMAP is construction of a spectroscopy-based catalog for systems with $M \gtrsim 10^{14} M_{\odot}$ and redshift $ z \leq 0.5$. Ultimately HectoMAP will be the basis for a cluster catalog based on a robust combination of photometric, X-ray, weak lensing, and spectroscopic methods. In fact, application of a friends-of-friends algorithm (e.g. \citealp{Huchra82, Ramella02, Robotham11, Tempel16}) applied to an earlier version of the entire HectoMAP catalog provided 166 friends-of-friends (FoF) systems for comparison with existing X-ray catalogs \citep{Sohn18b}. 

\citet{Sohn18a} use HectoMAP to test the redMaPPer (RM hereafter) catalog \citep{Rykoff14, Rykoff16}, a prototype for photometric cluster identification and cluster membership probabilities. In that study, the median number of spectroscopic HectoMAP members of the 104 RM clusters is $\sim 20$. Even at the lowest RM richness the fraction of real systems (purity) of the RM catalog is impressively high, $\sim90$\%. Figure \ref{cone} shows the 23 RM clusters in HectoMAP DR1 superimposed on the redshift survey cone diagram. The correspondence between the two catalogs is striking. In many cases the finger (elongation) in redshift space that corresponds to the cluster is apparent.

At the time of the \citet{Sohn18a} analysis the HectoMAP selection included a restriction on $(r - i)$ for galaxies with $r < 20.5$. This cut led to a systematic undersampling of clusters at redshift $z < 0.2$. Removal of this cut improved the sampling as expected; the enhanced observations add 114 ($\sim 9\%$) redshifts for RM cluster member candidates with $P_{mem} > 0$ ($P_{mem}$ is the radial and luminosity weighted RM membership probability \citep{Rykoff14}) in the HectoMAP DR1 region along with 25 new spectroscopic members not identified by RM.

\begin{figure*}
\centering
\includegraphics[scale=0.47]{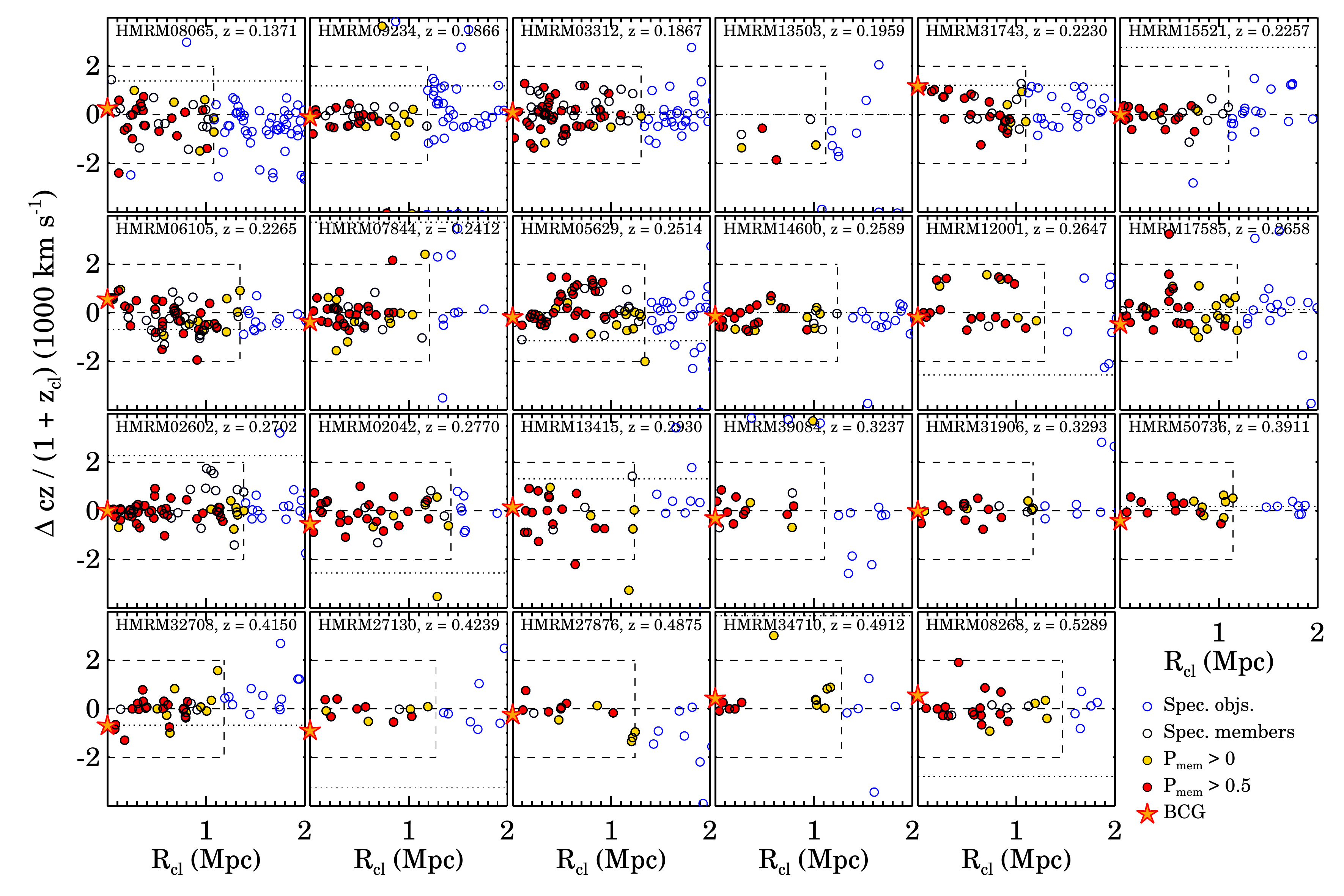}
\caption{R-v diagrams for redMaPPer cluster candidates in HectoMAP DR1.  
Red and yellow filled circles indicate member candidates with $P_{mem} \geq 0.5$ and $0.5 > P_{mem} > 0.0$, respectively,  identified by redMaPPer. 
Blue open circles show additional galaxies with spectroscopy at large projected radius. The orange star indicates the central galaxy identified by redMaPPer. The dashed box indicates the limits of member selection for the DR1-redMaPPer comparison in Table \ref{RMcomp}. Open circles inside this box indicate spectroscopic members identified by HectoMAP. The dashed horizontal line shows the mean spectroscopic redshift of the system and the dotted line shows the photometric redshift determined by redMaPPer. Note that the redMaPPer photometric redshift may lie outside the display box (see Table \ref{RMcomp}). }
\label{rv_rm}
\end{figure*}

\begin{figure*}
\centering
\includegraphics[scale=0.47]{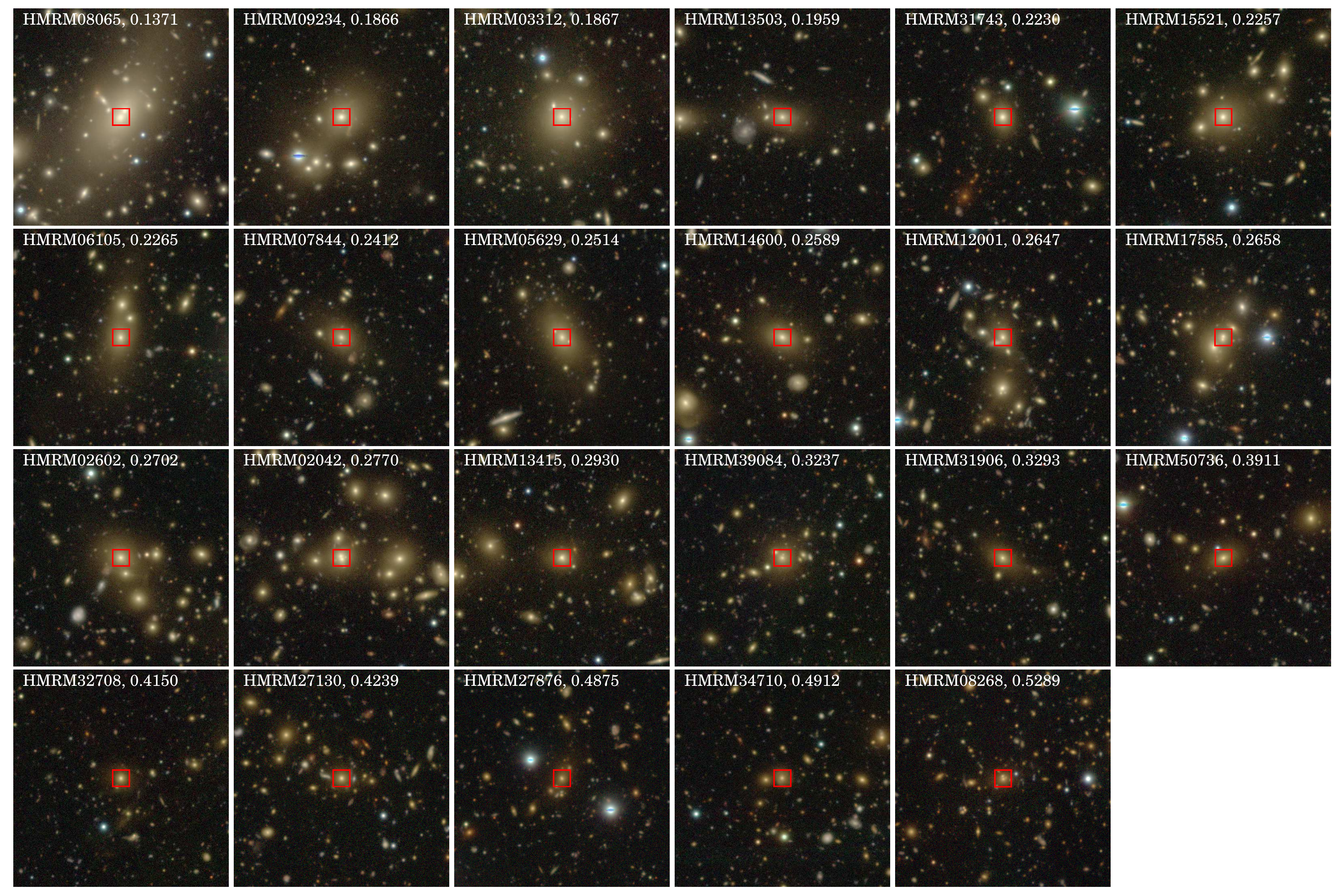}
\caption{Subaru/HSC images of the 23 redMaPPer clusters in HectoMAP DR1. Each image shows a $1.5\arcmin \times 1.5\arcmin$ field. A red square marks the central galaxy in each image. The labels indicate the redMaPPer ID and the redshift of the system. The color channels R, G, and B of the thumbnails are HSC-i, HSC-r, and HSC-g, respectively.}
\label{hsc_rm}
\end{figure*}

The $(r-i)$ selection remains an issue for one RM cluster candidate, HMRM13503 (Table \ref{RMcomp}). The typical spectroscopic survey completeness for RM member candidates with $(r-i) < 0.5$ and $P_{mem} > 0.5$ is $\sim 70\%$. However, the completeness for HMRM13503 is only $\sim 21\%$ because HMRM13503 is near the survey boundary. We have spectroscopic redshifts for only two member candidates with $P_{mem} > 0.5$. The low survey completeness prevents derivation of a spectroscopic redshift for this cluster candidate. Thus, we use the photometric redshift of the system from the redMaPPer catalog. We include HMRM13503 in Figures \ref{rv_rm} and \ref{hsc_rm}, but we limit further comments to the 22 more uniformly sampled systems. 

Figure \ref{rv_rm} shows phase space diagrams for all of the HectoMAP DR1 RM cluster candidates (ordered by redshift) with richness larger than 20. To derive the mean spectroscopic redshift for a system, we first select probable cluster members with $P_{mem} > 0.5$ that have spectroscopic redshifts. We compute the median redshift of these probable members and iterate by removing $3\sigma$ outliers. The median redshift of the remaining spectroscopically confirmed RM members is the cluster spectroscopic redshift. 

In Figure \ref{rv_rm}, red and yellow filled circles show member candidates identified by the RM algorithm: red circles indicate a RM membership probability $P_{mem} > 0.5$ and yellow circles indicate $0.5 > P_{mem} > 0$, respectively. In general, the RM candidate members with $P_{mem} > 0.5$ are closer to the cluster center (see also \citet{Sohn18a}). The photometric redshift from the RM catalog indicated by the dotted line is often offset from the spectroscopic redshift. The mean (median) offset is $1300~\kms$ ($1200~\kms$) and the dispersion in the offset is $3500~\kms$. This difference and scatter results from error in the photometric redshift ($\sim 3000~\kms$) and from the impact of foreground/background structures unresolved by RM.

The membership identification window finds members within a projected separation R$_{RM}$ and a rest-frame radial velocity difference $\Delta = |(c({z}_{galaxy}-{z}_{cl})/(1+{z}_{cl})| < 2000~\kms$, where R$_{RM}$ is a projected clustercentric radius, $z_{galaxy}$ is the spectroscopic redshift of the potential cluster member, and $z_{cl}$ is the spectroscopically determined cluster central redshift. We set the R$_{RM}$ limit to match the maximum R$_{RM}$ of the RM candidate members in each cluster \citep{Rykoff14}. This boundary tends to be larger for richer systems. We choose the relative rest-frame velocity limit by assessing the maximum range of spectroscopically identified members in known massive clusters (e.g., HeCS, \citealp{Rines13, Rines16, Rines18}). The resulting spectroscopic membership is insensitive to variations in $\Delta$ over the range $1500-2500~\kms$. Column 11 of Table \ref{hmdata} indicates spectroscopically determined RM cluster membership.
 
The RM candidate member list includes galaxies with $r < 22$. Thus HectoMAP with its limit of $r = 21.3$ does not include the apparently faintest candidates. HectoMAP also does not generally include objects bluer than $(r-i) = 0.5$ for $r > 20.5$. Table \ref{RMcomp} shows that the fractional completeness of the HectoMAP census of cluster member candidates generally decreases with the cluster redshift; HMRM13503 is an exception as a result of observational issues as discussed above. The number of cluster members identified by spectroscopy generally increases with the RM richness at a given redshift. The number of spectroscopically identified members ranges from 50-60 for richer systems (RM richness $\lambda \gtrsim 40$) at low redshift to 12-24 for systems of similar richness at redshift $z \gtrsim 0.3$.  At $z \gtrsim 0.3$ the HectoMAP sampling is often limited because most of the candidate members are fainter than the HectoMAP apparent magnitude limit. As expected, throughout the sample, a higher fraction of RM $P_{mem} > 0.5$ candidates are confirmed by spectroscopy. 

In a few cases apparently rich RM systems are  inflated by structure superposed along the line-of-sight (e.g., HMRM12001 where only $\sim 50\%$ of the RM $P_{mem} > 0$ candidates are spectroscopic members). At lower redshift, the broader color selection for galaxies with $r < 20.5$ in HectoMAP yields a larger sample of cluster members identified by spectroscopy but not included as RM candidate members. A median 32 RM members are confirmed by spectroscopy in the 22 RM clusters (column 6 of Table \ref{RMcomp}).

The RM central galaxies are obvious in the Subaru HSC images (Figure \ref{hsc_rm}). Each HSC image shows a $1.5^\prime \times 1.5^\prime$ region centered on the RM central. Four clusters, 31743, 06105, 07844, 08268,  contain a spectroscopic member brighter than the RM central in the SDSS r-band. The fraction of RM clusters with a galaxy brighter than the central is $18\%$, consistent with the estimates by \citet{Rykoff16} and \citet{Sohn18a}. In three of these systems, the brightest galaxy is significantly offset ($\sim 370$ kpc) from the RM center and the density of cluster members around the RM central is greater than around the brighter member. For the cluster HMRM 08268, the brightest galaxy is possibly a better choice for the cluster center.

\begin{deluxetable*}{lcccccccc}
\tablecaption{redMaPPer Clusters in HectoMAP DR1}
\tablecolumns{9}
\tabletypesize{\scriptsize}
\tablewidth{0pt}
\tablehead{
\colhead{ID} & \colhead{R.A.} & \colhead{Decl.} & 
\colhead{$N_{mem >0}^{1}$} & \colhead{$N_{mem > 0.5}^{2}$} & \colhead{$N_{spec, mem}^{3}$} & 
\colhead{$\lambda^{4}$} & \colhead{$z_{phot}$} & \colhead{$z_{spec}^{5}$}}
\startdata
HMRM08065 & 245.362203 &  42.761316 &  40/28/24 &  25/19/17 &  39 & $26.55 \pm  2.39$ & $0.1424 \pm 0.0045$ & 0.1371 \\
HMRM09234 & 246.598995 &  42.889029 &  52/49/24 &  24/24/16 &  33 & $24.92 \pm  2.46$ & $0.1913 \pm 0.0053$ & 0.1866 \\
HMRM03312 & 246.677190 &  42.669934 &  56/49/43 &  45/42/40 &  63 & $41.28 \pm  3.09$ & $0.1871 \pm 0.0050$ & 0.1867 \\
HMRM13503 & 244.826793 &  42.769710 &  38/13/ 4 &  24/ 5/ 2 &   6 & $23.37 \pm  2.20$ & $0.1959 \pm 0.0060$ & 0.1959 \\
HMRM31743 & 248.862917 &  43.163936 &  35/33/26 &  24/23/21 &  33 & $20.17 \pm  2.30$ & $0.2279 \pm 0.0080$ & 0.2230 \\
HMRM15521 & 246.874333 &  42.896084 &  37/31/18 &  25/19/16 &  26 & $24.46 \pm  2.64$ & $0.2371 \pm 0.0085$ & 0.2257 \\
HMRM06105 & 248.785849 &  42.800077 &  72/52/39 &  43/32/29 &  60 & $40.45 \pm  3.45$ & $0.2237 \pm 0.0068$ & 0.2265 \\
HMRM07844 & 245.321364 &  42.954352 &  58/50/34 &  32/31/25 &  40 & $31.25 \pm  2.64$ & $0.2566 \pm 0.0097$ & 0.2412 \\
HMRM05629 & 243.443971 &  43.268378 &  73/57/44 &  40/35/31 &  57 & $38.78 \pm  3.10$ & $0.2466 \pm 0.0083$ & 0.2514 \\
HMRM14600 & 242.339068 &  42.919114 &  58/43/24 &  28/22/17 &  28 & $28.29 \pm  2.90$ & $0.2590 \pm 0.0100$ & 0.2589 \\
HMRM12001 & 246.508451 &  43.704142 &  70/43/22 &  34/24/17 &  24 & $32.16 \pm  3.27$ & $0.2539 \pm 0.0094$ & 0.2647 \\
HMRM17585 & 245.464063 &  43.093565 &  56/44/34 &  26/24/19 &  36 & $27.05 \pm  2.86$ & $0.2664 \pm 0.0103$ & 0.2658 \\
HMRM02602 & 248.482328 &  43.091167 &  81/62/47 &  53/46/39 &  59 & $49.48 \pm  3.49$ & $0.2798 \pm 0.0106$ & 0.2702 \\
HMRM02042 & 247.063222 &  43.813437 &  98/66/30 &  52/41/24 &  34 & $49.29 \pm  3.37$ & $0.2660 \pm 0.0093$ & 0.2770 \\
HMRM13415 & 244.422275 &  42.539315 &  54/29/19 &  32/21/16 &  22 & $30.56 \pm  2.86$ & $0.2987 \pm 0.0125$ & 0.2930 \\
HMRM39084 & 248.935265 &  42.628005 &  64/27/12 &  22/11/10 &  14 & $23.19 \pm  2.83$ & $0.3769 \pm 0.0194$ & 0.3237 \\
HMRM31906 & 248.205718 &  43.390310 &  62/36/16 &  24/17/11 &  20 & $24.05 \pm  2.81$ & $0.3532 \pm 0.0184$ & 0.3293 \\
HMRM50736 & 247.060881 &  43.217691 &  57/26/18 &  20/12/11 &  18 & $24.58 \pm  3.45$ & $0.3919 \pm 0.0153$ & 0.3911 \\
HMRM32708 & 247.752305 &  43.033666 &  50/34/28 &  21/20/17 &  32 & $23.37 \pm  3.45$ & $0.4118 \pm 0.0147$ & 0.4150 \\
HMRM27130 & 249.234246 &  43.923393 &  52/23/12 &  26/10/ 8 &  13 & $27.46 \pm  3.38$ & $0.4086 \pm 0.0155$ & 0.4239 \\
HMRM27876 & 243.171435 &  43.416313 &  46/15/12 &  24/ 9/ 8 &  13 & $44.21 \pm  8.56$ & $0.5200 \pm 0.0168$ & 0.4875 \\
HMRM34710 & 243.046178 &  43.658707 &  48/22/12 &  17/ 8/ 6 &  12 & $38.05 \pm  7.60$ & $0.5102 \pm 0.0155$ & 0.4912 \\
HMRM08268 & 247.456116 &  43.839720 &  69/28/20 &  38/19/16 &  24 & $68.94 \pm 10.86$ & $0.5147 \pm 0.0142$ & 0.5289
\enddata 
\label{RMcomp}
\tablenotetext{1}{Number of member candidates / number of candidates with spectroscopy / spectroscopically identified members with $P_{mem} > 0$. }
\tablenotetext{2}{Number of member candidates / number of candidates with spectroscopy / spectroscopically identified members with $P_{mem} > 0.5$.}
\tablenotetext{3}{Number of spectroscopically identified members within $|\Delta cz / (1+z_{cl})| < 2000~\kms$.}
\tablenotetext{4}{Richness from the redMaPPer catalog}
\tablenotetext{5}{HectoMAP spectroscopic redshift of the system. }
\end{deluxetable*}

\begin{figure}
\centering
\includegraphics[scale=0.47]{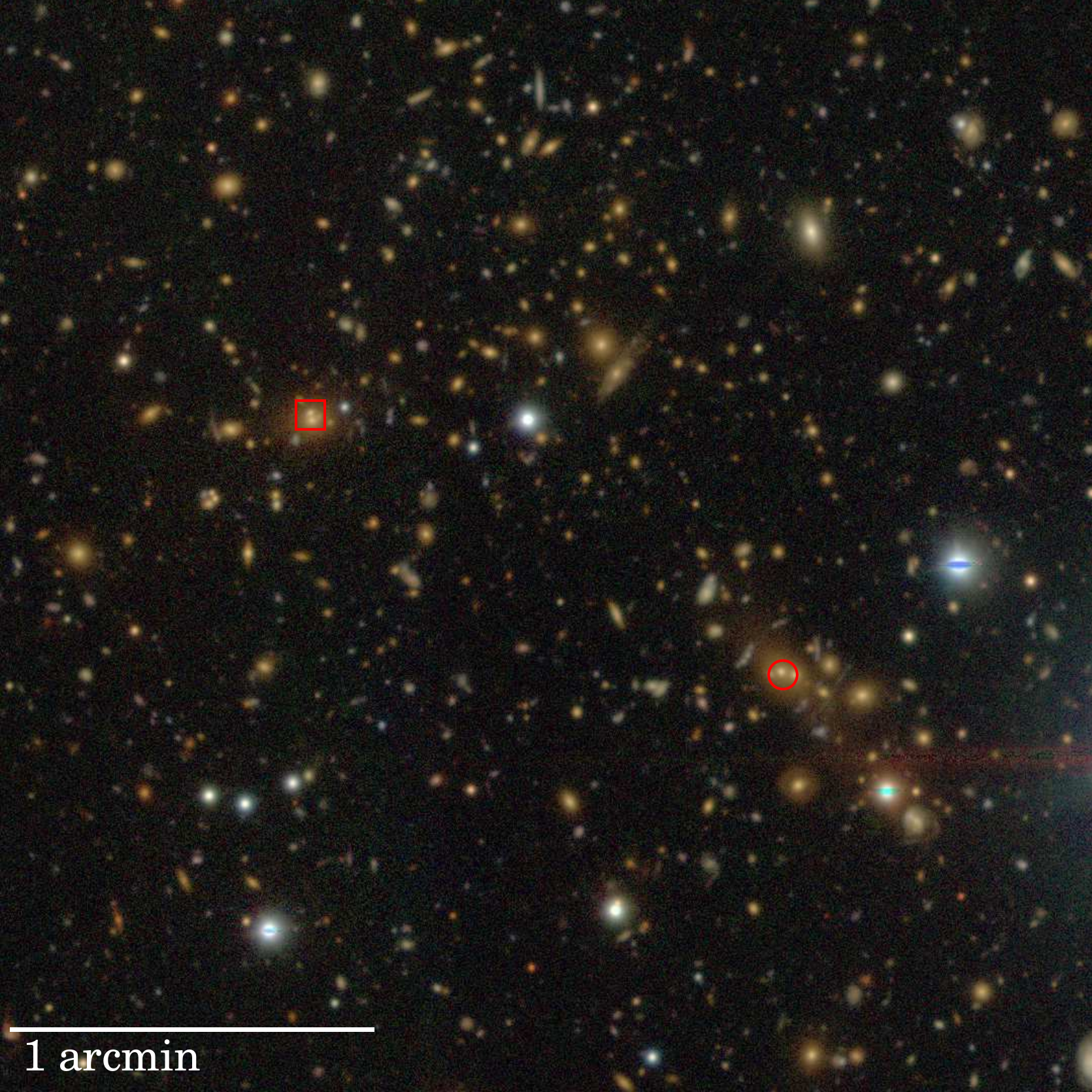}
\caption{Subaru/HSC image of HMRM 08268 ($z = 0.528$). The red square indicates the RM central galaxy and the red circle indicates the brightest member. Both the RM central and the brightest galaxy are associated with several strong lensing arc candidates. } 
\label{rm_uniq}
\end{figure}

Figure \ref{rm_uniq} displays the HSC image of HMRM 08268 (at $z = 0.528$). A red square marks the RM central and a red circle indicate the brightest spectroscopic member. Surprisingly, there are apparent strong lensing arcs associated with both the RM central and the brightest member. The projected separation between the two bright galaxies is $\sim 560$ kpc and their relative rest-frame velocity is $\sim 675~\kms$. In this system clustering around the brightest member is more impressive than it is around the RM central. HMRM 08268 is the subject of an intensive spectroscopic campaign with the MMT and Keck (Sohn et al. 2021); it is a candidate merging cluster.

The census of RM candidate systems in the HectoMAP DR1 confirms that these richness $\lambda > 20$ are bona fide clusters of galaxies. The spectroscopy provides important refinements of system membership by eliminating foreground/background contamination and by identifying additional members not included in the RM member candidate lists (Table \ref{RMcomp}). The survey also underscores the reliability of the RM central identification as an excellent proxy for the BCG; the most striking departure, HMRM 08268 is an interesting case that highlights the impact of spectroscopy in a comprehensive survey.

The RM systems are one element of the multi-technique cluster catalog enabled by HectoMAP. We plan to combine a redshift-based catalog, X-ray detections from e-ROSITA \citep{Finoguenov19, Sohn19}, and weak lensing detections from HSC data to construct this catalog of clusters with $z \lesssim 0.55$. The HSC imaging also makes a platform for identifying a set of strong lensing system candidates like HMRM 08268.

Stacked sets of clusters from this HectoMAP cluster catalog combined with lower redshift systems sampled well to large radius (e.g., \citealp{Rines06, Rines13, Rines16}) enable investigations of cluster growth. The redshift range encompasssed by HectoMAP is interesting because simulations and analytic models show that $z = 0$ clusters with masses of $10^{14-15}$ M$_{\odot}$ accreted half of their mass at an approximately constant rate between $z \sim 0.5$ and the current epoch \citep{vandenBosch14, Correa15, Pizzardo20}. \citet{Pizzardo20} develop a technique for direct measurement of the cluster accretion rate as a function of cluster mass and epoch. The small subset explored here predicts the rough number of spectroscopic members in stacked systems, the platform for a first direct dynamical test of the accretion history models in the HectoMAP redshift range.

\subsection{Test of HSC Photometric Redshifts}\label{Sec:photz}

Increasingly extensive photometric surveys make the use of photometric redshifts (hereafter $z_{phot}$s) imperative for studying galaxy evolution and for limiting the cosmological parameters. Larger and larger surveys with well-calibrated photometry in five or more bands have prompted more and more sophisticated development of $z_{phot}$ estimators. Template fitting along with more recent applications of machine learning provide platforms for analyzing the challenging current and future datasets. \citet{Tanaka18} also develop a hybrid technique they call FrankenZ that combines the strengths of template fitting and machine learning. 

\citet{Tanaka18} train a variety of $z_{phot}$ estimators with a large, careful compilation of 170k spectroscopic redshifts ($z_{spec}$) extending to $z \sim 4$. They then provide a variety of performance tests and demonstrate the efficacy of $z_{phot}$ over the redshift range $0.2 < z < 1.5$. Here we complement their investigation with HectoMAP DR1. The entire HectoMAP survey will provide a uniform sample of more than 110k redshifts for detailed tests of $z_{phot}$ for $z \lesssim 0.7$.

We compare the HectoMAP DR1 $z_{spec}$ with  HSC $z_{phot}$. We use $z_{phot}$s from the public HSC SSP DR2 catalog. The catalog provides only two of the $z_{phot}$ estimators explored by \citet{Tanaka18}: DeMP and Mizuki, a template-fitting code \citep{Tanaka15}. Because of its better performance \citep{Tanaka18}, we examine $\sim 110,000$ $z_{phot}$ computed with the DeMP code \citep{Hsieh14} for objects brighter than $r = 22$ in the HectoMAP DR1 region. The DeMP code fits each input galaxy based on a subset of the training set with photometry and colors closest to the target object. \citet{Tanaka18} use their dense training set to apply this regional polynomial fitting to the HSC data.

Among the HSC SSP public DR2 objects, 17,040 ($\sim 10\%$ of the number of galaxies in the \citet{Tanaka18} spectroscopic training sample) have a HectoMAP $z_{spec}$. The dense, relatively bright, red-selected HectoMAP sample complements the generally deeper datasets used by \citet{Tanaka18}. HectoMAP provides a test bed largely independent of the \citet{Tanaka18} training sets thus complementing their results.

\begin{figure}
\centering
\includegraphics[scale=0.5]{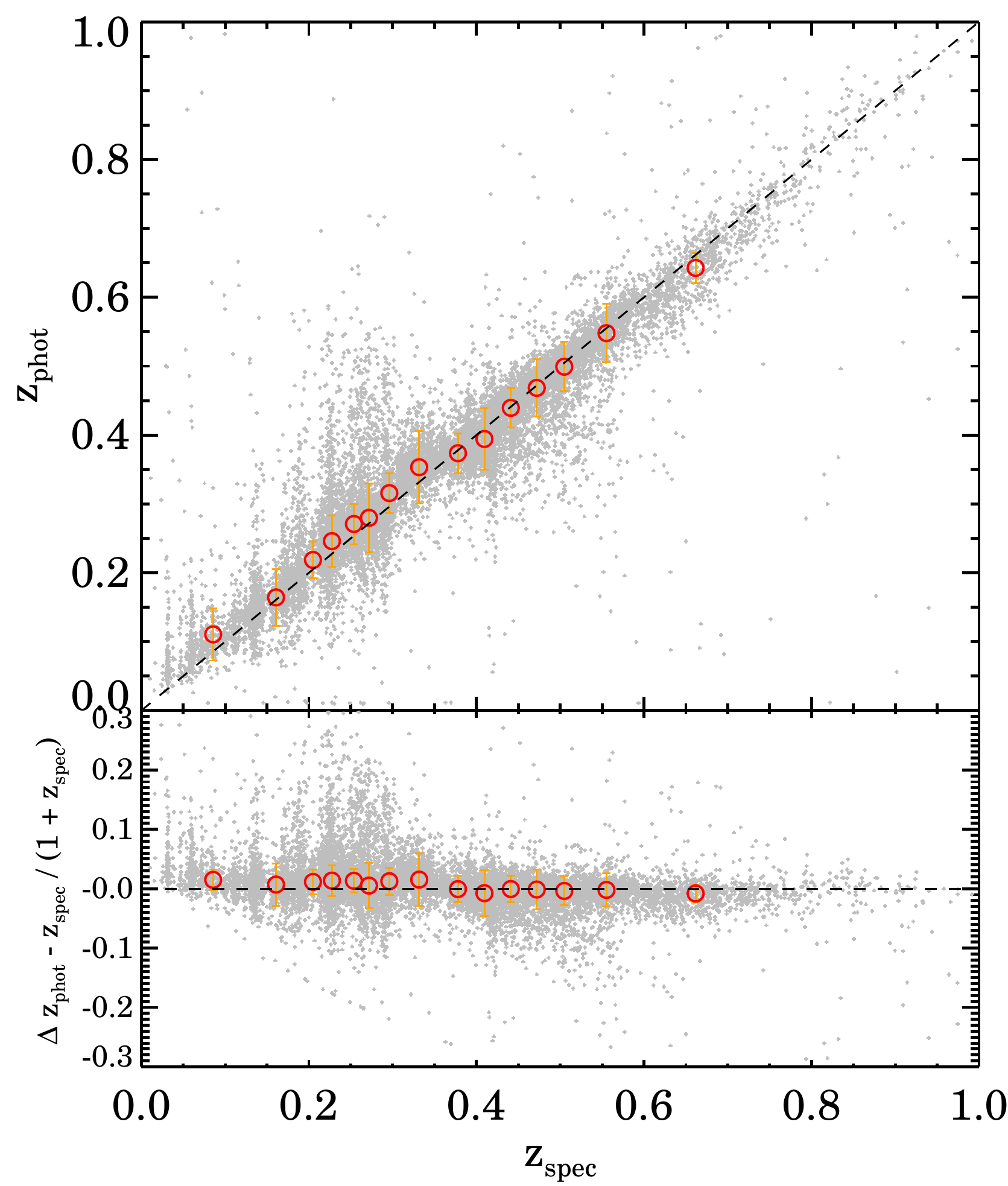}
\caption{(a) Comparison between spectroscopic redshifts and DeMP $z_{phot}$s from the HSC catalog in HectoMAP DR1. Red circles show the median of the distribution in each spectroscopic redshift bin. (b) Difference between $z_{spec}$ and $z_{phot}$ as a function of $z_{spec}$. }
\label{photz}
\end{figure}

Figure \ref{photz} (a) shows the DeMP $z_{phot}$ as a function of the HectoMAP $z_{spec}$. The two redshifts generally follow a one-to-one relation, but the scatter is large. Circles mark the median of the photo-z distribution in each $z_{spec}$ bin. There are small systematic shifts as a function of redshift. Extensions along the $z_{phot}$ direction occur where $z_{phot}$ differs substantially from $z_{spec}$.

Figure \ref{photz} (b) shows the difference between the photometric and spectroscopic redshifts as a function of spectroscopic redshift. The red symbols show the median redshift difference. In the $z_{spec}$ interval $0.2 - 0.4$ the median $z_{phot}$ exceeds $z_{spec}$. Overall, the median redshift difference is $0.004 \pm 0.047$. This typical redshift difference ($\sim 1100~\kms$) corresponds to the line-of-sight velocity dispersion of massive galaxy clusters.

\begin{figure}
\centering
\includegraphics[scale=0.5]{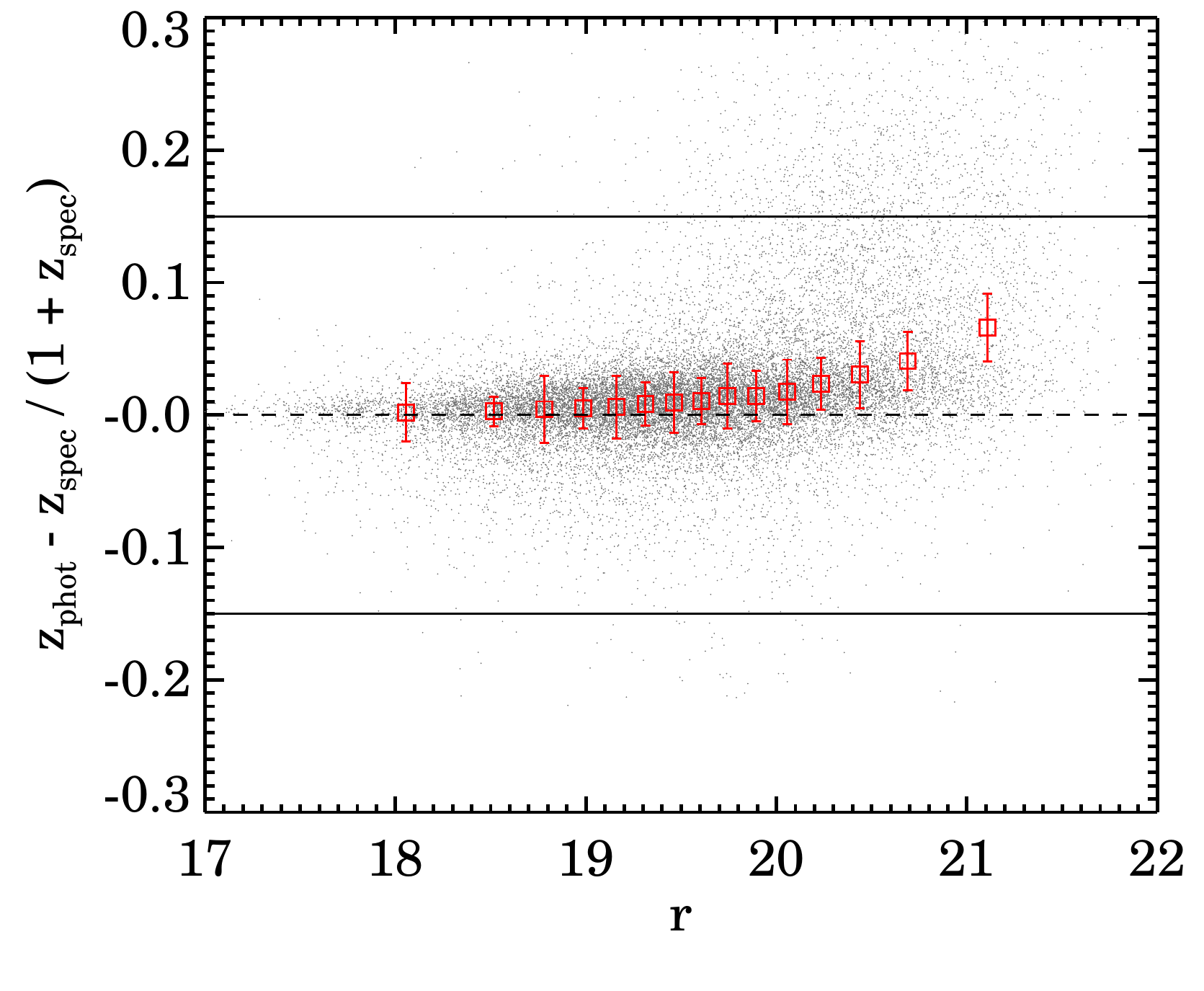}
\caption{Difference between photometric and spectroscopic redshifts as a function of $r_{petro,0}$  for HectoMAP DR1 galaxies with $0.2 < z_{spec} < 0.3$ (gray points). 
The red symbols show the median and $1\sigma$ standard deviation. }
\label{photz_mag}
\end{figure}

The accuracy of the $z_{phot}$ measurements is a strong function of apparent magnitude. Figure \ref{photz_mag} shows the difference between photometric and spectroscopic redshifts as a function of $r-$band magnitude for galaxies in the spectroscopic redshift range $0.2 < z_{spec} < 0.3$. The red squares indicate the median difference. The redshift difference increases at fainter magnitude ($r > 20$ mag). Furthermore, the number of outliers ($|\Delta z| > 0.15$) is also high at $r > 20$. A similar test at higher redshift requires a survey to a fainter magnitude limit.

\begin{figure*}
\centering
\includegraphics[scale=0.32]{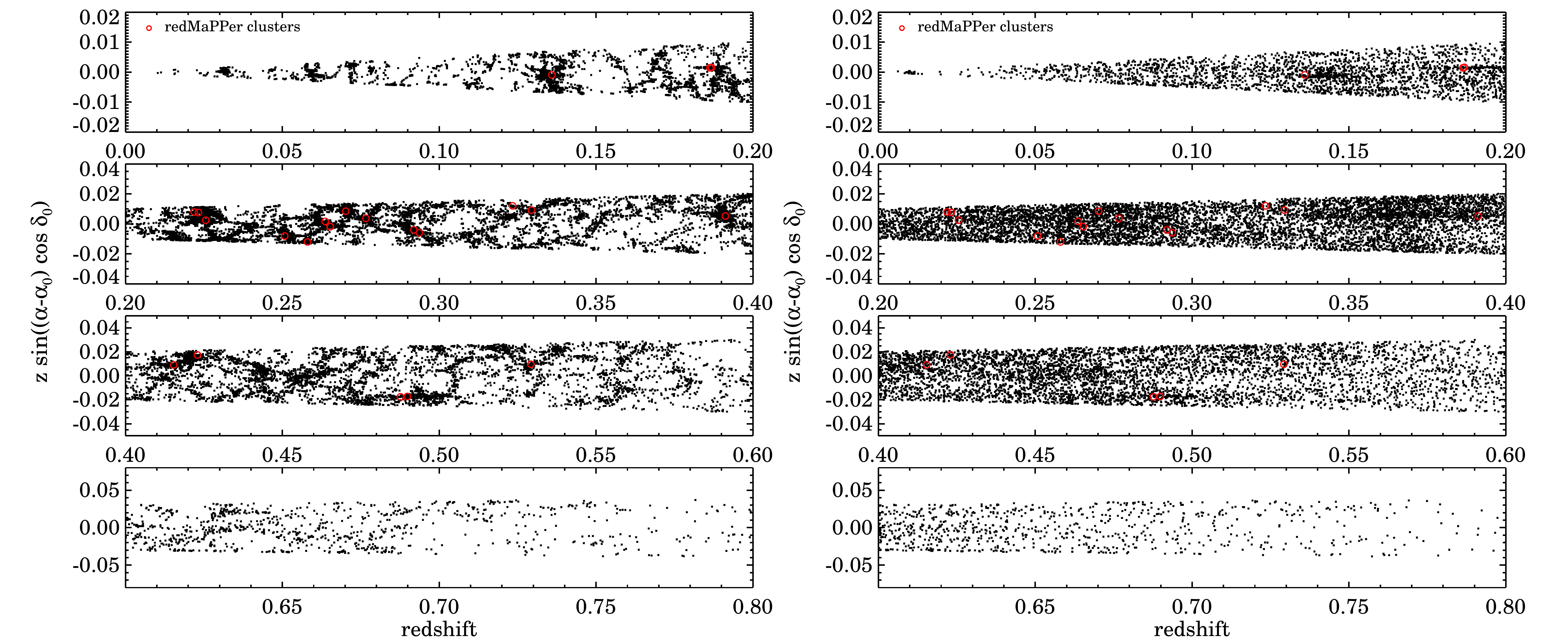}
\caption{Cone diagrams for  HectoMAP DR1  based on spectroscopic redshifts (left) and photometric redshifts (right).} 
\label{cone_phot}
\end{figure*}

Figure \ref{cone_phot} is a visual demonstration of the impact of $z_{phot}$ on the large-scale structure. For ease of direct comparison, the left-hand panel of the Figure shows the cone diagrams of Figure \ref{cone}. The points indicate individual galaxies and the red dots show the positions of RM clusters in redshift space. The right-hand panel shows the cone diagrams based on DeMP $z_{phot}$. The error in a typical DeMP $z_{phot}$ is $0.05 \pm 0.12$ or, equivalently, $\sim 15000~\kms$. Overall large-scale density gradients are approximately preserved, but the filaments, walls and voids characteristic of the large-scale structure of the universe become invisible.

\citet{Tanaka18} use a variety of metrics to test their $z_{phot}$ algorithms (see also \citealp{Pasquet19}). For comparison with their analysis, we compute three metrics for the HectoMAP DR1 sample: the bias, the dispersion, and the loss rate.

\begin{figure*}
\centering
\includegraphics[scale=0.5]{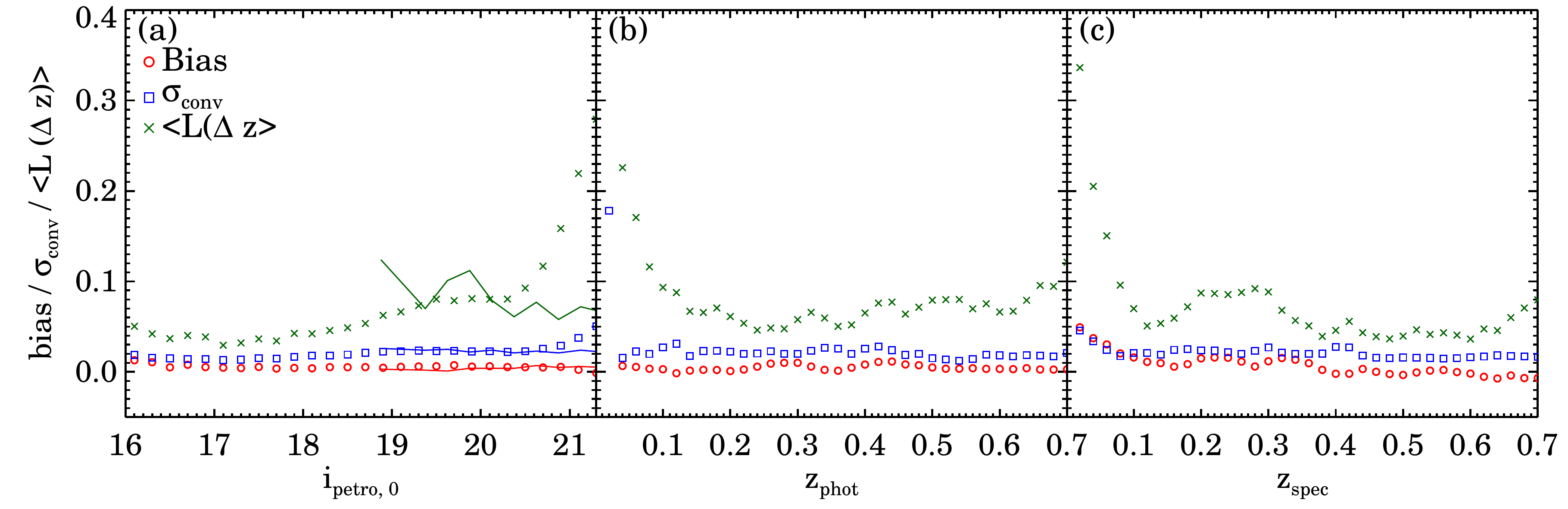}
\caption {(Left) Bias (red points), $\sigma_{conv}$ (blue points), and $<~L(\Delta z)~>$ (green points) of $z_{phot, DeMP}$ as a function of SDSS $i-$band magnitude. The correspondingly colored lines show comparable results from \citet{Tanaka18}.
(Middle) Same as the left panel but as a function of $z_{phot}$. (Right) Same as the left panel, but as a function of $z_{spec}$.}
\label{photz_test}
\end{figure*}

The bias measures the systematic offset, if any, between the photometric and spectroscopic redshifts. As in \citet{Tanaka18} we compute
\begin{equation}
\Delta z =  (z_{phot} - z_{spec}) / (1 + z_{spec}). 
\end{equation}
Figure \ref{photz_test} shows the median value of this offset as a function of $i$, $z_{phot}$, and $z_{spec}$ (red points). Lines in the left-hand panel of Figure \ref{photz_test} show the analogous results of \citet{Tanaka18} (their Table 2). For this measure, the agreement between the HectoMAP DR1 test and the \citet{Tanaka18} testbed is remarkable. There is essentially no dependence of the bias on either the apparent magnitude or the photo-z (central panel of Figure \ref{photz_test}). In contrast with \citet{Tanaka18}, we also plot the metrics as a function of $z_{spec}$ (right-hand panel of Figure \ref{photz_test}). Over the redshift range we sample, the bias  changes systematically with $z_{spec}$; at lower redshifts the photometric estimate overestimates the redshift and at the highest redshift, the $z_{phot}$ is an underestimate. This behavior is also evident in Figure \ref{photz}.

Next we compute the conventional dispersion
\begin{equation}
\sigma_{conv} = 1.48\ {\rm MAD}(\Delta z).
\end{equation}
The MAD $(\Delta z)$ is the median absolute $\Delta z$. Blue points in Figure \ref{photz_test} show the result. In the overlapping apparent magnitude range, the agreement with \citet{Tanaka18} is excellent. The dispersion increases slightly for fainter apparent magnitudes as expected. The large dispersion at low spectroscopic redshift in the right-hand panel may result from the inclusion of more blue, emission-line objects at this redshift. At larger redshift, the HectoMAP DR1 test sample is dominated by absorption-line objects as a result of the red selection.

\citet{Tanaka18} develop a metric they call the loss function. This metric has the  advantage that combines traditional metrics in a single function. The loss function is a continuous version of the outlier fraction that incorporates the impact of both the bias and dispersion. The expression for the loss function is
\begin{equation}
L(\Delta z) = 1 - { 1 \over {1 + (\Delta z/\gamma)^2}},
\end{equation}
where \citet{Tanaka18} adopt $\gamma = 0.15$ to reflect the standard limit $\Delta z = 0.15$ in the calculation of the outlier fraction.

Figure \ref{photz_test} shows the loss function for HectoMAP DR1 (green points) compared with the \citet{Tanaka18} result (green line). For $i$-band magnitudes between 19 and 20 the results agree. At fainter magnitudes the loss function for HectoMAP exceeds the \citet{Tanaka18} result. This difference is driven primarily by faint galaxies with spectroscopic redshifts in the range $0.2 < z_{spec} < 0.3$ (Figure \ref{photz_mag}). These galaxies have relatively low absolute luminosities and some have low surface brightness.

The dependence of the loss function on photometric and spectroscopic redshift (center and right-hand panels of Figure \ref{photz_test}) provides further insight into the subtleties of the relative behavior of the two measures. The loss function increases at $z_{phot} < 0.1$.  It rises slowly from a $z_{phot} = 0.2$ to $z_{phot} = 0.7$ reflecting the large scatter visible in Figures \ref{photz} and \ref{photz_mag}. The scatter around the one-to-one correspondence between photometric and spectroscopic redshifts in Figure \ref{photz} is most pronounced in the redshift range $0.2 < z_{spec} < 0.3$. In the central panel of Figure \ref{photz_test} this scatter spreads the data over a wide range of $z_{phot}$; in the right-hand panel where we display the dependence of the loss function of $z_{spec}$, there is an obvious positive bump in the loss function confined to this range. The cleaner dependence of the loss function on spectroscopic redshift highlights the impact of the larger photo-z scatter and bias for faint objects in a relatively narrow redshift range (Figure \ref{photz_mag}).

Exploration of the HSC SSP $z_{phot}$ based on HectoMAP DR1 shows both the power and subtle limitations of the $z_{phot}$. The HectoMAP analysis confirms and amplifies the results of \citet{Tanaka18} for bright magnitudes and low redshifts. Comparison of spectroscopic and photometric redshifts as a function of spectroscopic redshift reveals a systematic bias that is a function of redshift; this bias does not appear either as a function of apparent magnitude or as a function of photometric redshift.

Investigation of other metrics, particularly the loss function, as a function of $z_{spec}$ shows that at least for red objects, the $z_{phot}$ perform well down to a redshift $z = 0.15$, in slight contrast $z = 0.2$ limit stated by \citet{Tanaka18}. The HectoMAP study complements \citet{Tanaka18} by highlighting the impact of the increased bias and scatter of $z_{phot}$ for faint objects in a fixed spectroscopic redshift range. The full HectoMAP sample will enable this kind of detailed investigation over the redshift range 0.1 - 0.7.

\section{Conclusion}

HectoMAP is a dense red-selected redshift survey covering 54.64 square degrees to a limiting $r = 21.3$ in a narrow strip across the northern sky. The complete survey will include $\sim 110,000$ redshifts. The first data release (HectoMAP DR1) covers 8.7 square degrees. HectoMAP DR1 includes 17,313 galaxy redshifts along with stellar mass and $\dn$ for nearly all of the galaxies. Among these galaxies, we acquired 14,767 redshifts with the Hectospec wide-field fiber instrument on the MMT.

The HSC SSP survey \citep{Aihara18} covers the entire HectoMAP region; HectoMAP DR1 encompassses the smaller area HSC SSP DR1 region. The combination of deep photometry and dense spectroscopy enables investigations that combine both strong and weak lensing with spectroscopy. The excellent seeing over large portions of the HSC SSP data  also enables detailed investigations of the size evolution of the quiescent galaxy population.

We outline the quality and completeness of the HectoMAP DR1 data with an eye toward demonstrating its applications to problems in cosmology and large-scale structure. We emphasize the power of a dense survey in the study of the definition and evolution of voids and massive clusters over the redshift range 0.2 to 0.7.

As a demonstration of the applicability of HectoMAP to astrophysical and technical issues in large-scale structure we revisit the properties of redMaPPer clusters. The typical number of spectroscopic members of the 22 uniformly surveyed clusters is $\sim 31$. The HectoMAP observations combined with the HSC SSP photometry highlight a fascinating merging system, HMRM08268 at $z = 0.528$. This system contains two luminous galaxies associated with strong lensing arcs. HMRM08268 is already the subject of a Keck telescope campaign.

We also highlight the insights HectoMAP provides on HSC SSP photometric redshifts. Dense coverage of the redshift range uncovers a subtle systematic bias in the median photometric redshift as a function of spectroscopic redshift. The HectoMAP data also show that in a fixed spectroscopic redshift range, 0.2 to 0.3, both the bias and the scatter in photometric relative to spectroscopic redshifts increase significantly for fainter objects. The full HectoMAP survey will provide similarly detailed tests for redshifts from 0.2 to 0.7.

HectoMAP currently fills a niche between large volume surveys at lower redshift and surveys that reach to much greater redshift. Over the next few years a remarkable array of multi-object spectrographs on large facilities will provide dense surveys with much larger areal coverage in the HectoMAP range and beyond. HectoMAP enables initial explorations of many astrophysical issues that can inform the design of these future surveys.

\appendix
We identify 524 stars in the HectoMAP DR1 region based on spectroscopy. These objects are classified as galaxies in the SDSS photometric catalog, but their absolute radial velocities are  $< 730~\kms$. Table \ref{Hstar} lists these stars including the SDSS object ID, R.A., Decl, and the redshift (or blueshift) and its uncertainty. 

\begin{deluxetable}{lccc}\label{Hstar}
\tablecaption{Spectroscopically Identified Stars in  HectoMAP DR1}
\tablecolumns{9}
\tabletypesize{\scriptsize}
\tablewidth{0pt}
\tablehead{\colhead{ID} & \colhead{R.A.} & \colhead{Decl.} & \colhead{z}}
\startdata
1237651250442732340 & 247.644228 &  43.681899 & $-0.00007 \pm 0.00015$ \\
1237651250442731853 & 247.668384 &  43.773074 & $-0.00020 \pm 0.00018$ \\
1237651250442732438 & 247.578860 &  43.610764 & $-0.00065 \pm 0.00014$ \\
1237651250442732509 & 247.633911 &  43.609529 & $-0.00015 \pm 0.00028$ \\
1237655347284213786 & 244.386122 &  42.925257 & $-0.00011 \pm 0.00001$
\enddata
\end{deluxetable}

\acknowledgments
We thank Perry Berlind and Michael Calkins for operating Hectospec. We also thank Susan Tokarz and Jaehyon Rhee for their important contributions to the data reduction. We thank Antonaldo Diaferio, Ken Rines, and Scott Kenyon for discussions that clarified the paper. This paper uses data products produced by the OIR Telescope Data Center, supported by the Smithsonian Astrophysical Observatory. J.S. gratefully acknowledges the support of a CfA Fellowship. The Smithsonian Institution supported the research of M.J.G., D.F., and S.M.M. This research has made use of NASA’s Astrophysics Data System Bibliographic Services. 

Funding for the Sloan Digital Sky Survey IV has been provided by the Alfred P. Sloan Foundation, the U.S. Department of Energy Office of Science, and the Participating Institutions. SDSS-IV acknowledges support and resources from the Center for High Performance Computing at the University of Utah. The SDSS website is www.sdss.org. SDSS-IV is managed by the Astrophysical Research Consortium for the Participating Institutions of the SDSS Collaboration including the Brazilian Participation Group, the Carnegie Institution for Science, Carnegie Mellon University, Center for Astrophysics | Harvard \& Smithsonian, the Chilean Participation Group, the French Participation Group, Instituto de Astrof\'isica de Canarias, The Johns Hopkins University, Kavli Institute for the Physics and Mathematics of the Universe (IPMU) / University of Tokyo, the Korean Participation Group, Lawrence Berkeley National Laboratory, Leibniz Institut f\"ur Astrophysik Potsdam (AIP), Max-Planck-Institut f\"ur Astronomie (MPIA Heidelberg), Max-Planck-Institut f\"ur Astrophysik (MPA Garching), Max-Planck-Institut f\"ur Extraterrestrische Physik (MPE), National Astronomical Observatories of China, New Mexico State University, New York University, University of Notre Dame, Observat\'ario Nacional / MCTI, The Ohio State University, Pennsylvania State University, Shanghai Astronomical Observatory, United Kingdom Participation Group, Universidad Nacional Aut\'onoma de M\'exico, University of Arizona, University of Colorado Boulder, University of Oxford, University of Portsmouth, University of Utah, University of Virginia, University of Washington, University of Wisconsin, Vanderbilt University, and Yale University.

The Hyper Suprime-Cam (HSC) collaboration includes the astronomical communities of Japan and Taiwan, and Princeton University. The HSC instrumentation and software were developed by the National Astronomical Observatory of Japan (NAOJ), the Kavli Institute for the Physics and Mathematics of the Universe (Kavli IPMU), the University of Tokyo, the High Energy Accelerator Research Organization (KEK), the Academia Sinica Institute for Astronomy and Astrophysics in Taiwan (ASIAA), and Princeton University. Funding was contributed by the FIRST program from the Japanese Cabinet Office, the Ministry of Education, Culture, Sports, Science and Technology (MEXT), the Japan Society for the Promotion of Science (JSPS), Japan Science and Technology Agency (JST), the Toray Science Foundation, NAOJ, Kavli IPMU, KEK, ASIAA, and Princeton University. This paper makes use of software developed for the Large Synoptic Survey Telescope. We thank the LSST Project for making their code available as free software at http://dm.lsst.org. This paper is based [in part] on data collected at the Subaru Telescope and retrieved from the HSC data archive system, which is operated by Subaru Telescope and Astronomy Data Center (ADC) at National Astronomical Observatory of Japan. Data analysis was in part carried out with the cooperation of Center for Computational Astrophysics (CfCA), National Astronomical Observatory of Japan.

\facilities{MMT Hectospec, Subaru Hyper Suprime Cam}

\bibliography{ms}

\end{document}